\documentclass[11pt,sort&compress]{elsarticle}
\usepackage[paper=a4paper,marginratio={1:1,5:6}]{geometry}
\makeatletter
\def\ps@pprintTitle{%
 \let\@oddhead\@empty
 \let\@evenhead\@empty
 \def\@oddfoot{\centerline{\thepage}}%
 \let\@evenfoot\@oddfoot}
\makeatother
\usepackage{natbib}
\usepackage[hyperref]{xcolor}
\definecolor{darkgreen}{rgb}{0.01, 0.75, 0.24}
\definecolor{darkblue}{HTML}{2B66D3}
\usepackage[colorlinks=true,
            linkcolor=darkblue,
            urlcolor=darkblue,
            citecolor=darkblue,linkcolor=darkblue]{hyperref}      

\usepackage{amsmath}
\usepackage{amsfonts}
\usepackage{siunitx}
\usepackage{slashed}
\usepackage{accents}
\usepackage{amssymb}
\usepackage{mathrsfs}
\usepackage{mathtools}
\usepackage[utf8]{inputenc}
\usepackage[T1]{fontenc}
\let\oldbibliography\thebibliography
\renewcommand{\thebibliography}[1]{%
  \oldbibliography{#1}%
  \setlength{\itemsep}{1.4pt}%
}

\usepackage[cal=boondox,calscaled=1]{mathalfa}
\DeclareMathAlphabet{\bbvar}{U}{BOONDOX-ds}{m}{n}
\DeclareMathAlphabet{\mathsl}{\encodingdefault}{\familydefault}{m}{sl}

\usepackage[defaultsans,scale=0.9]{opensans}
\usepackage{psfrag}
\usepackage{pstool}
\usepackage{caption}
\usepackage{tabularx}
\usepackage{multirow}
\usepackage{pbox}

\usepackage{graphicx}
\usepackage{subcaption}
\newcommand{\hook}{\text{\large{$\lrcorner$}}}
\usepackage{relsize}

\newcommand{\qq}[1]{``#1''} 

\newcommand{\di}{\mathrm{d}}
\usepackage{tensor}
\newcommand{\ou}[3]{\tensor{#1}{^{#2}_{#3}}}
\newcommand{\uo}[3]{\tensor{#1}{_{#2}^{#3}}}

\newcommand{\I}{\mathrm{i}} 
\newcommand{\E}{\mathrm{e}} 
\newcommand{\C}{\mathbb{C}}

\newcommand{\R}{\mathbb{R}}

\newcommand{\eref}[1]{(\ref{#1})}

\DeclareMathAlphabet{\bbgreek}{U}{bbold}{m}{n}

\newcommand{\mtext}[1]{\text{\it #1}}
\usepackage{tikz-cd}
\usetikzlibrary{arrows.meta, automata, 
                positioning, 
                quotes}
\usepackage[low-sup]{subdepth}

\newcommand\vpm{\mathbin{\vcenter{\hbox{
  \oalign{\hfil$\scriptstyle+$\hfil\cr
          \noalign{\kern-.3ex}
          $\scriptscriptstyle({-})$\cr}}}}}
\DeclareMathAlphabet{\sfit}{OT1}{\sfdefault}{m}{it}
\DeclareMathAlphabet{\sfbfit}{OT1}{\sfdefault}{sb}{it}
\DeclareMathAlphabet{\mathsf}{OT1}{\sfdefault}{sb}{n}

\definecolor{darkgreen}{rgb}{0.01, 0.75, 0.24}
\definecolor{darkblue}{HTML}{2B66D3}

\usepackage[multiple, flushmargin]{footmisc}
\let\originalleft\left
\let\originalright\right
\renewcommand{\left}{\mathopen{}\mathclose\bgroup\originalleft}
\renewcommand{\right}{\aftergroup\egroup\originalright}

\usepackage{scalerel}

\newcommand{\dbarvar}{{\mathrm{d}\mkern-7.5mu\lower.18ex\hbox{$\textasciitilde$}\mkern-1.5mu}}

\renewcommand{\emph}[1]{{\it #1}}

\begin{document}

\begin{abstract}
There has been recently renewed interest in the quantisation of gravity by considering local subsystems on light-like hypersurfaces. The main purpose of this paper is to present a theory-independent perspective on these developments assuming only basic knowledge of quantum theory and general relativity.
In addition, we present a top-down approach to constructing local amplitudes in causal diamonds. The fundamental building block is a slab of light-like geometry (e.g.\ a segment of a light cone embedded into spacetime). Each slab is a three-dimensional light-like hypersurface bounded by two cuts, its past and future corners. After briefly reviewing the timeless formalism of quantum theory, we equip each null slab a with a kinematical Hilbert space that factorizes into constituents associated to the three-dimensional interior and its two corners. 
Upon assuming the existence of vacuum states for the bulk and boundary symmetries and a fundamental projector onto physical states, we explain how to introduce local amplitudes by contracting boundary states according to the pattern of a causal diamond. Finally, we show that the resulting local transition amplitudes satisfy Ward identities and charge conservation for the underlying symmetries. The paper closes with a summary and discussion for how different approaches to quantum gravity can realise our proposal in practice.

  \end{abstract}%
\title{Foundational Structure of Local Amplitudes in Quantum Gravity}
\author{Charalampos Theofilis and Wolfgang Wieland}
\address{Institute for Quantum Gravity, Theoretical Physics III, Department of Physics\\Friedrich-Alexander-Universität Erlangen-Nürnberg\\Staudstraße 7, 91052 Erlangen. Germany
\\{\vspace{0.5em}\normalfont \today}}

\maketitle
\vspace{-1.5em}
\hypersetup{
  linkcolor=black,
  urlcolor=black,
  citecolor=black
}
{\tableofcontents}\vspace{-0.5em}
\hypersetup{
  linkcolor=darkblue,
  urlcolor=darkblue,
  citecolor=darkblue,
}
\begin{center}{\noindent\rule{\linewidth}{0.4pt}}\end{center}\newpage
\section{Introduction and Motivation}
\noindent Quantum gravity is often presented as a mere formal-algorithmic problem. The goal is to complete the Dirac programme for general relativity and find a quantisation of its Poisson algebra of observables \cite{PhysRev.160.1113,ashtekar,kiefer, thiemann, rovelli,Dittrich:2004cb}. The most recent incarnation of this bottom-up approach is based on the  Hamiltonian analy\-sis of general relativity in causal domains, which are subregions of spacetime that are bounded by light-like (null) surfaces. By now, there is a wide literature with many related developments \cite{Reisenberger:2012zq,Fuchs:2017jyk,Reisenberger:2018xkn,Wieland:2017cmf,Wieland:2025qgx,Wieland:2017zkf,AndradeeSilva:2022iic,Jacobson:2022gmo,Bub:2024nan,Freidel:2023bnj,Wieland:2024dop,Wieland:2025LP,Ciambelli:2024swv,Wieland:2020gno,Fiorucci:2025twa,Donnay:2022aba,Klinger:2025hjp}. The present article lays out a complementary top-down perspective on the same problem. The main purpose is to present a theoretical framework for how such an approach could function in practice. The paper assumes only basic knowledge of general relativity, gauge theories, and quantum theory. Our goal is to present the general structure and core ideas to a broad audience in quantum gravity, quantum foundations and quantum information science.\smallskip

Let us briefly explain the basic terminology and key ideas.  The proposal is based on a quasi-local viewpoint \cite{Wang:2008jy,Szabados:2004vb}, in which we consider local subsystems of the gravitational field \cite{Donnelly:2016auv,Giddings:2019hjc,Carrozza:2022xut,DeVuyst:2024fxc} as the primitive elements of the theory to begin with. These subsystems describe the physical degrees of freedom in a compact spacetime domain. Such gravitational subsystems are necessarily open, rather than closed, because they can interact with the outside environment \cite{Kabel:2022efn,Wieland:2020gno}.  The opposite viewpoint is that there is a global wave function of the universe \cite{Hartle:1983ai} from which all local physics follows through e.g.\ a consistent histories approach to quantum theory \cite{Griffiths:1984aa}. At the classical level, such a global state  is characterized by the initial data on an initial surface (Cauchy surface) for fixed asymptotic boundary conditions (asymptotically flat, de Sitter or anti-de Sitter). Such a viewpoint on quantum gravity seems problematic to us, because it is at odds with the probabilistic nature of quantum theory. In quantum theory, we obtain probabilities for measurement outcomes. These probabilities provide the best possible fit to the correlations and observed frequencies that we measure in repeated runs of the same physical experiment. 
Since we only have access to local measurements, the operational necessity of a universal wave function of the universe is unclear. From a global perspective, the primary elements are the amplitudes between asymptotic boundary states. Asymptotic boundary states represent different asymptotic initial data, different worlds or superpositions thereof. It is quite unclear what the operational meaning of such asymptotic amplitudes would be in a theory of quantum gravity.  There is no ensemble available to local observers in spacetime that consists of repeated measurement outcomes for asymptotic boundary states. Realistic experiments are local and happen within a finite spacetime domain.
Any such local quantum process consists of the preparation of the experiment and the actual measurement. The entire process is confined to a causal diamond which is the intersection of the past and future Cauchy development of measurement and preparation. The boundary of the causal diamond
is a light-like boundary, where we can freely prescribe initial data. Quantum gravity should then predict the probabilities that we can measure in an ensemble of such regions, threaded along the observer's worldline.\footnote{See also \cite{Schlick1948-SCHGKU-2} for a similar viewpoint.}  

Besides these operational considerations there are also a number of very practical technical reasons why null surfaces can simplify the problem \cite{ChoquetBruhat:2010ih,Chrusciel:2012ap,Chrusciel:2022ozm,Aretakis:2021lzi}. 
At the classical level, the phase space is the space of the solutions to the field equations. The state of solutions can be coordinatized by the initial data on a Cauchy surface (assuming global hyperbolicity). Cauchy surfaces can be spacelike or null. Further, they do not need to be smooth. Working on null hypersurfaces rather than spacelike hypersurfaces is useful, because there is  more structure available that we can utilise to build physical observables: null hypersurfaces are generated by light rays that provide a natural ordering structure, a sort of scaffolding that simplifies the canonical analysis.\smallskip 

In addition, the constraint algebra simplifies on null hypersurfaces \cite{Reisenberger:2012zq,Fuchs:2017jyk,Reisenberger:2018xkn}. This can be understood at a rather elementary level without going into too much mathematical details. Constraints arise from the pull-back of the field equations to the initial hypersurface.\footnote{In what follows, we skip most of the mathematical details and focus on the conceptual sides of the argument. The pull-back is a way to restrict co-tensorfields (e.g.\ the metric) to submanifolds in spacetime. What is important to note is that by taking the pull-back of the field equations to the initial hypersurface, we only obtain equations that involve derivatives that are intrinsic (tangential) to the hypersurface. In this way, we obtain  constraint equations, rather than evolution equations, which would contain transversal (time) derivatives that take us away from the initial surface.} The pull-back of the Einstein three-form $\ast F_{\alpha\beta}\wedge e^\beta$ is a top-form on the initial hypersurface.\footnote{In here, $e^\alpha$ is the co-tetrad that diagonalises the metric $ds^2 =\eta_{\alpha\beta}e^\alpha\otimes e^\beta$, with $\eta_{\alpha\beta}$ denoting the flat Minkowski metric. In addition, $\ou{F}{\alpha}{\beta}=\di\ou{\omega}{\alpha}{\beta}+\ou{\omega}{\alpha}{\gamma}\wedge\ou{\omega}{\gamma}{\beta}$ is the curvature two-form of the spin connection $\ou{\omega}{\alpha}{\beta}$ and $\ast \ou{F}{\alpha}{\beta}$ denotes the Hodge dual of the curvature two-form, see e.g.\ \cite{Krasnov_2020} for further details.} The vacuum Einstein equations set this top form to zero. For each point on the initial hypersurface, the free Lorentz index $\alpha = 0, 1, 2,3$ will then label four independent constraints. On a space-like hypersurface, these are the vector and Hamiltonian constraints, which are the generators of hypersurface deformations (see \hyperref[fig1]{Figure 1} below). On a null hypersurface, only three of them are generators of gauge symmetries, i.e.\ first-class constraints \cite{HennauxTeitelboim_book}. There is a simple and intuitive reason for why this is so:
\begin{figure}[h]
\centering
\includegraphics[width=0.8\textwidth]{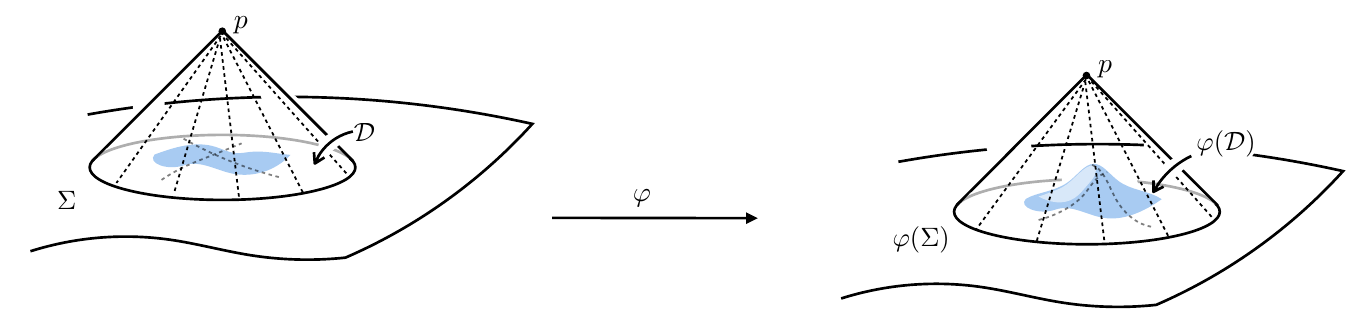}
\caption{A diffeomorphism $\varphi$, which becomes the identity outside of a small (four-dimensional) neighbourhood that surrounds a compact domain $\mathcal{D}\subset\Sigma$. The diffeomorphism maps a spatial Cauchy surface $\Sigma$ into a new surface $\varphi(\Sigma)$. For an observer placed in the causal future (past) of $\mathcal{D}$, the initial data at $\Sigma$ and $\varphi(\Sigma)$ characterise the same physical state. In a background-invariant theory, the two configurations are gauge equivalent (indistinguishable by local experiments at $p$).\label{fig1}}
\end{figure}%
 Take a small diffeomorphism $\varphi\in\mathrm{Diff}(\mathcal{M}:\mathcal{M})$ of compact support acting on a solution of Einstein's equations $(\mathcal{M}, ds^2)$. Consider then  a local subsystem on a portion $\mathcal{N}$ of a null hypersurface (\hyperref[fig1]{\color{darkblue}{Figure 2}} below). Choose the diffeomorphism $\varphi$ in such a way that it deforms only a small compact domain of 
$\mathcal{N}\subset\mathcal{M}$ and maps $\mathcal{N}$ into a new surface $\varphi(\mathcal{N})$, which is no longer null. Can such a diffeomorphism be possibly generated by the Hamiltonian vector field of a gauge constraint on the phase space of the subsystem? This can not possibly
be so, for otherwise we could imagine the situation depicted in \hyperref[fig2]{\color{darkblue}{Figure 2}}. 
\begin{figure}[h]
\centering
\includegraphics[width=0.7\textwidth]{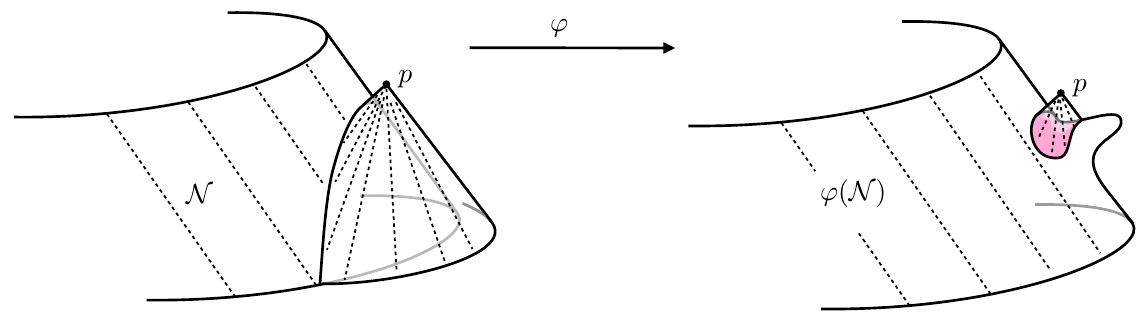}
\caption{The diffeomorphism $\varphi$ deforms a three-dimensional null hypersurface $\mathcal{N}\subset\mathcal{M}$ into a new hypersurface $\varphi(\mathcal{N})\subset\mathcal{M}$ that is no longer null. No matter how small this diffeomorphism is, the new surface $\varphi(\mathcal{N})$ contains data that is not available on $\mathcal{N}$. Therefore, the motion on phase space that takes initial data on $\mathcal{N}$ to $\varphi(\mathcal{N})$ cannot be a classical canonical transformation (unitary transformation in quantum theory).\label{fig2}}
\end{figure}%
On the left, the data on $\mathcal{N}$ is not enough to predict what happens at $p$. On the right, there is a small neighbourhood of $\varphi(\mathcal{N})$, whose future Cauchy development contains $p$. Notice then that the Hamiltonian vector field of a gauge constraint can not explore new data on phase space. All points on the gauge orbit represent the same physical state. Yet, there is more initial data on $\varphi(\mathcal{N})$ than there is on $\mathcal{N}$. Thus, there is a contradiction. A more detailed canonical analysis confirms this expectation \cite{Reisenberger:2012zq,Fuchs:2017jyk,Reisenberger:2018xkn,PhysRevD.91.064043}.

 Thus, there are less gauge transformations on a null surface than there are on a spacelike hypersurface. In short, we get rid of the scalar (Hamiltonian) constraint. One of the many difficulties with imposing the scalar constraint at the quantum level is that it does not generate a Lie algebra, because there are structure functions rather than structure constants \cite{Bojowald:2016hgh,Teitelboim1973542,adm,Hojman197688,PhysRev.160.1113}. On a null hypersurface, we have again a Lie algebra of constraints.\smallskip

In addition, there are  further simplifications that happen when considering the Hamiltonian analysis on null hypersurfaces. \emph{First}, the constraint equations are simple transport equations for canonical variables along the null generators, rather than elliptic equations on spatial hypersurfaces \cite{Isenberg:2013iva}.  
These constraints, which are the pull-back of the Einstein equations to the null hypersurface, can be solved on a hierarchy of phase spaces. The smallest of these phase spaces only deals with the Raychaudhuri constraint \cite{Fuchs:2017jyk,Reisenberger:2018xkn,Reisenberger:2012zq,Wieland:2025qgx,Wieland:2021vef,Ciambelli:2023mir}, which is the projection of the Einstein equations into the direction of the null generators. 
\emph{Second}, fields on different light rays commute under the Poisson bracket \cite{Fuchs:2017jyk,Reisenberger:2018xkn,Reisenberger:2012zq,Wieland:2025qgx,Wieland:2021vef,Ciambelli:2023mir}. In addition, each light ray carries a gauge symmetry: reparametrisations of the null coordinate, which are generated by the integral of the Raychaudhuri constraint smeared against appropriate test functions. Conformal field theories (CFTs) in two dimensions carry representations of such one-dimensional diffeomorphisms.  The whole problem of how to quantise gravitational null initial data can be then mapped into an equivalent problem for an auxiliary CFT on each light ray \cite{Wieland:2025qgx,Ciambelli:2024swv}. The quantisation of this CFT provides a natural candidate for a quantum theory of gravitational null initial data \cite{Wieland:2025qgx} compatible with earlier results in loop quantum gravity \cite{Rovelliarea,Haggard:2023tnj,AshtekarLewandowskiArea,FernandoBarbero:2009ai}. The main purpose of this paper is to explain the general mathematical structure and conceptual foundations for such an approach and how it allows to construct quasi-local amplitudes from the symmetries of the boundary states.\smallskip

The paper is divided into three parts. The next two sections are basic reviews about the timeless formulation of quantum mechanics (\hyperref[sec2]{\color{darkblue}Section 2}) and the Noether theo\-rem and covariant phase space formalism (\hyperref[sec3]{\color{darkblue}Section 3}). The main point of the paper is outlined in \hyperref[sec4]{\color{darkblue}Section 4}, in which we explain how to construct quasi-local transition amplitudes in causal domains from the states and intertwiners that for the boundary symmetry algebra.

\section{Timeless Quantum Mechanics}\label{sec2}
\noindent Quantum theory admits a timeless formulation on a kinematical boundary Hilbert space $\sfit{K}$ that treats measurements and state preparations in a time-symmetric manner \cite{PhysRevD.27.2885,PhysRevD.42.2638,Rovelli:2009ee,rovelli}. A field theoretic realisation of this framework is based on earlier developments on topological quantum field theories and quantum gravity in three spacetime dimensions \cite{Atiyah:1989vu,Witten:1988hc,Carlip:1998uc}. 
The formalism is tailored to background independent approaches to quantum gravity, in which there is no external time \cite{Atiyah:1989vu,Witten:1988hc,Carlip:1998uc}. The underlying state space is a  boundary Hilbert space that describes the joint boundary data at the closed boundary of a compact spacetime domain \cite{PhysRevD.42.2638,Rovelli:2009ee,rovelli,Barrett:1995mg,Oeckl2003,Oeckl:2005bv,Oeckl:2012ni,Oeckl:2016tlj}.\smallskip

At the classical level, the corresponding boundary value problem is over-determined. The  constraints that remove the unphysical configurations from the kinematical boundary phase space can be inferred from the Hamilton--Jacobi equations, whose solutions impose constraints between the past and future components of the boundary phase space.\footnote{If $(p_f,q_f;p_i,q_i)$ are canonical coordinates on the boundary phase space, which splits into initial $(p_i,q_i)$ and final configurations $(p_f,q_f)$, the constraints are given by $p_f=(\partial_{q_f}S)(q_i,q_f)$ and $p_i=-(\partial_{q_i}S)(q_i,q_f)$, where $S(q_i,q_f)$ is a solution to the Hamilton--Jacobi equation.}\smallskip 
 
  A simple way to formally construct the elements of the kinematical boundary Hilbert space is to use a functional Schrödinger representation \cite{Kiefer:1991xy,Oeckl:2012qb} in which the states are complex-valued functionals $\Psi[\varphi|_{\partial\mathcal{M}}]$ of the field configurations at the boundary.
 Let us briefly review this formalism in the ordinary quantum mechanical setting, where we deal with one-dimensional time intervals rather than four-dimensional spacetime regions. 
In this simple case, the boundary splits into two disconnected parts, and we have 
\begin{equation}
\sfit{K}=\mathcal{K}_+\otimes \mathcal{K}_-.
\end{equation}
Elements of the two Hilbert space factors represent the measurement and preparation that define an idealized quantum process. This Hilbert space is called \emph{kinematical}, because there are no correlations at this stage between future and past events. The transition amplitudes, which enforce such correlations, are introduced only in a secondary step. 
\smallskip 

The two factors are related by the Hilbert space inner product:\footnote{A more general framework in which the two Hilbert space factors can have different dimensions has been studied in the context of quantum Regge calculus, see \cite{Hohn:2014uvt}. } $\mathcal{K}_-$ is the dual Hilbert space of $\mathcal{K}_+$, i.e.\ $\mathcal{K}_-=\mathcal{K}_+^\ast$. Consider then a generic element of $\sfit{K}$, which will be entangled between its past and future components,
\begin{equation}
\Psi=\sum_{i}\langle\psi_{i}|\otimes|\phi_{i}\rangle\in\sfit{K}.\label{psi-state}
\end{equation}
For any such state, there is a natural pairing, given by the linear map $\left(\emptyset\right|:\sfit{K}\rightarrow\C$,
\begin{equation}
\left(\emptyset|\Psi\right\rangle:=\sum_{i}\langle\psi_{i}|\phi_{i}\rangle.
\end{equation}
This map defines a state $(\emptyset|$ that lies in the algebraic dual space $\sfit{K}'$ of $\sfit{K}$. If $\sfit{K}$ is infinite-dimensional, this state is not normalizable.\smallskip
 
The transition amplitudes are defined in terms of a generalized projector \cite{rovelli, Perez:2012wv}, which is a linear map $P$ on $\mathcal{K}_\pm$ that can map $\mathcal{K}_\pm$ into possibly non-normalizable states (elements of the double algebraic dual of $\mathcal{K}_\pm$). In all relevant cases, this projector, which maps kinematical states into physical states, can be understood  as the kernel of a set of constraints $C_\alpha$ that generate the gauge symmetries of the theory. Schematically
\begin{equation}
\forall \alpha:C_\alpha P=P C_\alpha=0,\qquad[C_\alpha,C_\beta]=\mathcal{O}(C),
\end{equation}
for some unspecified labels $\alpha\in I$ that distinguish the individual gauge generators. At the classical level, the constraints become functions on phase space that weakly commute under the Poisson bracket. In  general relativity, these constraints are the Hamiltonian generators of the hypersurface deformation algebra \cite{Teitelboim1973542,adm,Hojman197688,PhysRev.160.1113}.\smallskip

The dynamics is encoded into the transition amplitudes, which are the matrix elements of the projector onto physical states. For a generic kinematical boundary state \eref{psi-state}, we obtain 
\begin{equation}
A(\Psi)=\left(\emptyset|(P\otimes\bbvar{1})\Psi\right\rangle=\left(\emptyset|(\bbvar{1}\otimes{P})\Psi\right\rangle=\sum_{i}\langle\psi_{i}|P\phi_{i}\rangle.
\end{equation}
Notice that the amplitudes define a new state $(0|\in\sfit{K}'$, i.e.\ the covariant vacuum \cite{rovelli}, which is an element of the algebraic dual $\sfit{K}'$ of the boundary Hilbert space $\sfit{K}$,
\begin{equation}
A(\Psi)=:(0|\Psi\rangle.
\end{equation}

Given the amplitudes, we infer probabilities for quantum transitions.  If $\{E_i^+\}_{i\in I}$ and $\{E^-_j\}_{j\in J}$ are two pairwise commuting POVMs (positive operator-valued probability measures) on $\sfit{K}$, we can use Born's rule to obtain the probability measure
\begin{equation}
p(E_i^+|E_j^-)=\frac{\left(0\right|E_i^+E^-_j\left|0\right)}{\left(0\right|E^-_j\left|0\right)}\label{Brule}.
\end{equation}
The POVMs $\{E_i^+\}_{i\in I}$ are in one-to-one correspondence with measurement outcomes (quantum events) $\{e_i\}_{i\in I}$. Following a statistical interpretation of quantum mechanics, the probabilities \eref{Brule} provide the best possible fit to the frequencies that we observe for the occurrences of the random events $\{e_i\}_{i\in I}$ in repeated runs of the same experiment (process). Each process is modelled by a choice of boundary state $E^-_j$. This boundary state generalizes the ordinary notion of density matrices on $\mathcal{K}_\pm$. Each $E^-_j$ is an element of $\sfit{K}\otimes\sfit{K}'$ and defines a so-called process matrix $W(\cdot):=\left(0\right|(\cdot)E^-_j\left|0\right)/\left(0\right|E^-_j\left|0\right)$, which allows for entanglement between the future and past components of the boundary \cite{Oreshkov:2012aa,Araujo:2015qya,HardyOTQT:2012}.\smallskip

{\it Example:} Consider a system coupled to a in ideal clock. The corresponding kinematical Hilbert space is $\mathcal{K}_{+}\otimes\mathcal{K}_-$, where $\mathcal{K}_-^\ast=\mathcal{K}_{+}=L^2(\R^2,\di t\,\di x)$. In this Hilbert space, there are kinematical states that are peaked around some average values of time and space. The dynamics is imposed by solving the Schrödinger equation
\begin{align}
{C}=\I\hbar\partial_t-H_o(\hat{p},\hat{x}).\label{SchrdngrConst}
\end{align}
Solutions to the constraint are the time average of kinematical states. The projector onto the solution space, which is the space of physical states, is given by
\begin{equation}
P=\int\di t\,\E^{-\frac{\I}{\hbar}t{C}}=2\pi\hbar\,\delta({C}).
\end{equation}
 Given a kinematical state $\Psi_o\in\mathcal{K}_\pm$ that is peaked around some configuration $(t_o,x_o)$, the corresponding physical state $P\Psi_o$ will be peaked around an entire trajectory $(t,x(t))$ with initial conditions $x(t_o)=x_o$. This makes it intuitively clear, why such states are typically non-normalizable.\footnote{An exception occurs for systems in which the trajectories are closed.} 
Consider then the following product states
\begin{align}
|\Psi_{\pm}\rangle&=|t_{\pm}\rangle\otimes|\psi_{\pm}\rangle,\label{bndrystate}\\
\langle t,x|\Psi_\pm\rangle&\equiv\Psi_{\pm}(t,x)=\delta(t_\pm-t)\,\psi_\pm(x),
\end{align}
The corresponding covariant boundary state is
\begin{equation}
\Psi=\langle\Psi_{+}|\otimes|\Psi_{-}\rangle.
\end{equation}
All time dependence is now encoded into the choice of boundary states. The transition amplitudes $A(\Psi)$ for the boundary states \eref{bndrystate} are the matrix elements of the usual unitary time evolution operator
\begin{equation}
W(\Psi)=\langle\Psi_{+}|P\Psi_{-}\rangle=\langle\psi_+|\E^{-\frac{\I}{\hbar}(t_+-t_-)H_o}\psi_-\rangle.
\end{equation}

This timeless reformulation of the standard quantum mechanical formalism is relevant for quantum gravity, because it clarifies the mathematical structure a potential quantum theory of gravity can have \cite{rovelli}. The practical necessity for such a timeless formulation is an immediate consequence of the equivalence principle. The equivalence principle implies general covariance and general covariance implies that there is no preferred time coordinate. Time is conjugate to energy. As there are no preferred clocks, there is also no preferred Hamiltonian. It would be highly unexpected therefore to have a quantum theory of gravity in which there is a unique Schrödinger type of evolution equation to calculate time-dependent transition amplitudes $A(\psi_-\rightarrow\psi_+,\Delta t)=\langle\psi_+|\E^{-\frac{\I}{\hbar}\Delta t H_o}\psi_-\rangle$. Instead, we seek for a more general framework, in which we only have timeless amplitudes $A(\Psi)$ for generic boundary states $\Psi\in\sfit{K}$. The duration $\Delta t$ of a quantum process, which is characterized by state preparation and measurement, can then only be inferred from the choice of boundary states. In that sense, time itself becomes quantum.

\section{Covariant Phase Space, Charges, Noether's Theorem}\label{sec3}
\noindent To clarify our presentation, we briefly review here in this section some basic aspects of Noether's theorem, edge modes and the covariant phase space formalism \cite{Ashtekar:1990gc,Wald:1999wa,Harlow:2020aa,Wieland:2017zkf,Oliveri:2019gvm,Wieland:2020gno,Freidel:2023bnj} that are relevant for our construction. In the next section, we take the basic concepts to the quantum level and propose how to build, in principle, local amplitudes solely by imposing gauge invariance at the boundary of a causal diamond. 
At the classical level, the underlying Ward identities return the Noether flux-balance laws considered below.\smallskip

For definiteness, consider here a local and covariant Lagrangian for some $p$-form field $\varphi^{\mathfrak{a}}\in \Omega^p(\mathcal{M}:\mathbb{V})$, where $\mathbb{V}$ is some unspecified target space and $\mathfrak{a},\mathfrak{b},\mathfrak{c},\dots$ are internal or (co-)tangent space indices for the various components of the $\mathbb{V}$-valued $p$-form fields.\footnote{Such Lagrangians include the case of  vacuum general relativity and general relativity coupled to the standard model of particle physics \cite{Rezende:2009sv,thiemann,Krasnov_2020}.} 
The Lagrangian $L[\varphi^{\mathfrak{a}},\di\varphi^{\mathfrak{a}}]$, which is a four-form on spacetime, is a local functional of $\varphi^{\mathfrak{a}}\in \Omega^p(\mathcal{M}:\mathbb{V})$ and its \emph{kinetic momentum}, i.e.\ the $(p+1)$-form $\pi^\mathfrak{a}=\di\varphi^{\mathfrak{a}}$. We assume a theory with a covariant Lagrangian. This is to say that for all four-dimensional diffeomorphisms $f\in\mathrm{Diff}(\mathcal{M}:\mathcal{M})$, field configurations $\varphi^{\mathfrak{a}}\in\Omega^p(\mathcal{M}:\mathbb{V})$  and kinetic momenta $\pi^{\mathfrak{a}}\in\Omega^{p+1}(\mathcal{M}:\mathbb{V})$ and $x\in\mathcal{M}$, the following identity is satisfied,
\begin{equation}
\big(f^\ast L[\varphi^{\mathfrak{a}},\pi^{\mathfrak{a}}]\big)(x)=L[f^\ast\varphi^{\mathfrak{a}},f^\ast\pi^{\mathfrak{a}}](x).\label{inv-actn}
\end{equation}
The Lagrangian is differentiable, if the chain rule holds. For all linearized field configurations $\delta\varphi^{\mathfrak{a}},\delta\pi^{\mathfrak{a}}$, i.e.\ tangent vectors $\delta$ on field space, we must then have
\begin{equation}
\delta L[\varphi^{\mathfrak{a}},\pi^{\mathfrak{a}}](x)=\delta\phi^{\mathfrak{b}}(x)\wedge\frac{\partial L[\varphi^{\mathfrak{a}},\pi^{\mathfrak{a}}](x)}{\partial\varphi^{\mathfrak{b}}}+\delta\pi^{\mathfrak{b}}(x)\wedge\frac{\partial L[\varphi^{\mathfrak{a}},\pi^{\mathfrak{a}}](x)}{\partial\pi^{\mathfrak{b}}}.
\end{equation}
This in turn allows to split the variation of the Lagrangian into the Euler--Lagrange equations and a boundary term, which determines the pre-symplectic potential
\begin{equation}
 \delta[{L}]\big|_{\pi^{\mathfrak{a}}=\di\phi^{\mathfrak{a}}}={E}(\delta)+\di(\vartheta(\delta)),
\end{equation}
where
\begin{align}
E(\delta)&=\delta\phi^{\mathfrak{b}}(x)\wedge\left(\frac{\partial L[\varphi^{\mathfrak{a}},\pi^{\mathfrak{a}}](x)}{\partial\varphi^{\mathfrak{b}}}-(-1)^p\di\frac{\partial L[\varphi^{\mathfrak{a}},\pi^{\mathfrak{a}}](x)}{\partial\pi^{\mathfrak{b}}}\right)\bigg|_{\pi^{\mathfrak{a}}=\di\varphi^{\mathfrak{a}}},\\
\vartheta(\delta)&=\delta\phi^{\mathfrak{b}}(x)\wedge\frac{\partial L[\varphi^{\mathfrak{a}},\pi^{\mathfrak{a}}](x)}{\partial\varphi^{\mathfrak{b}}}\bigg|_{\pi^{\mathfrak{a}}=\di\varphi^{\mathfrak{a}}}.
\end{align}
Through Noether's theorem, the invariance of the Lagrangian under four-dimensional diffeomorphisms \eref{inv-actn} implies a conservation law: consider a smooth one-parameter family of diffeomorphisms $f_\varepsilon\in\mathrm{Diff}(\mathcal{M}:\mathcal{M})$. For any fixed $x\in\mathcal{M}$, the derivative $\frac{\di}{\di\varepsilon}\big|_{\varepsilon=0}f_\varepsilon(x)=:\xi_x[f]$ defines a tangent vector $\xi_x^a\in T_x\mathcal{M}$. This definition naturally extends to any $p$-form field $\varphi^\mathfrak{a}\in\Omega^p(\mathcal{M}:\mathbb{V})$: if $f^\ast_\varepsilon$ is the pull-back, the Lie derivative is given by $\mathcal{L}_\xi(\cdot)=\frac{\di}{\di\varepsilon}\big|_{\varepsilon=0}f^\ast_\varepsilon(\cdot)$. Using the basic properties of the Lie derivative,\footnote{The Lie derivative of a $p$-form can be expressed in terms of the exterior derivative \qq{$\di$} and the interior product \qq{$\xi\hook$} as $\mathcal{L}_\xi(\cdot)=\di(\xi\hook\cdot)+\xi\hook(\di\cdot)$.} it is then straight forward to see
\begin{equation}
\mathcal{L}_\xi(L)=\di(\xi\hook L)=E(\mathcal{L}_\xi)+\di(\vartheta(\mathcal{L}_\xi)).
\end{equation}
If the equations of motion are satisfied, we obtain a conserved current 
\begin{equation}
 j_\xi=\vartheta(\mathcal{L}_\xi)-\xi\hook L,
\end{equation}
i.e.\ $\di j_\xi=0$. At least locally, there exists then by Poincaré's lemma  a two-form $q_\xi$, i.e.\ the charge aspect, such that $j_\xi=\di q_\xi$.\smallskip

In what follows, our fundamental building block is what we call a null slab of geometry, see \hyperref[fig3]{\color{darkblue}Figure 3} below. It consists of a three-dimensional null surface $\mathcal{N}$, i.e.\ the bulk geometry, which is bounded by two disjoint spatial hypersurfaces $\mathcal{C}_+$ and $\mathcal{C}_-$ (corners), such that $\partial\mathcal{N}=\mathcal{C}_+\cup\mathcal{C}_-^{-1}$. Given the Noether current $j_\xi$ and charge aspect $q_\xi$, we define the flux
\begin{align}
H_\xi[\mathcal{N}]&=\int_{\mathcal{N}}j_\xi.
\end{align}
In the same way, we define the boundary charges
\begin{align}
Q_\xi[\mathcal{C}]&=\oint_{\mathcal{C}}q_\xi.
\end{align}
Noether's theorem implies the flux balance law
\begin{equation}
Q_\xi[\mathcal{C}_+]-Q_\xi[\mathcal{C}_-]=H_\xi[\mathcal{N}].\label{charge-flx}
\end{equation}
At the quantum level, the flux-balance laws turn into constraints that imply non-trivial entanglement  between edge modes located at $\mathcal{C}_\pm$ and the radiative data on $\mathcal{N}$.\smallskip

In here, we assumed $\xi^a$ to be a smooth vector field on $\mathcal{N}$. From now on and to include unspecified gauge couplings, we will be more generic and assume that the gauge parameter $\xi^a$ takes values in some unspecified boundary symmetry algebra $\mathfrak{g}_{\mathcal{N}}$. The restriction to the two corners defines the corresponding corner symmetry algebra $\mathfrak{g}_{\mathcal{C}_\pm}\ni\xi_\pm=\xi\big|_{\mathcal{C}_\pm}$, which are dual to edge modes of the boundary symmetries \cite{Balachandran:1994up,PhysRevD.51.632,DonnFreid,Donnelly:2016rvo,Wieland:2017zkf,Wieland:2017cmf,Wieland:2021vef,Gomes:2016mwl,Speranza:2017gxd,Geiller:2017xad,Takayanagi:2019tvn,Francois:2021aa,Freidel:2019ees,Freidel:2020xyx,PhysRevLett.128.171302,Freidel:2021cjp,Donnelly:2020xgu,Goeller:2022rsx,Giesel:2024xtb}
Concrete realisations of these corner symmetry charges, which are dual to classical and quantum reference frames \cite{Loveridge2018,Giacomini:2017zju,Giacomini:2019fvi,Vanrietvelde:2018pgb,Hoehn:2019owq,Castro-Ruiz:2021vnq} at the two corners, can be found in \cite{Giesel:2024xtb,Wieland:2017zkf,Wieland:2020gno,Cresto:2024fhd,Freidel:2020xyx,Cresto:2024mne,Donnay:2024qwq,Neri:2025fsh}.

\section{Top-down Approach for Local Amplitudes in Causal Diamonds}\label{sec4}

\subsection{Kinematical boundary Hilbert space}	
\noindent In the following, we assume that the kinematical Hilbert space at the null boundary has the following tensor product structure:
 \begin{equation}
 \sfit{K}_{\mathcal{N}}= \mathcal{K}_{\mathcal{C}_+}\otimes\mathcal{K}_{\mathcal{N}}\otimes\mathcal{K}_{\mathcal{C}_-}.
 \end{equation}
In here, $\mathcal{K}_{\mathcal{C}_\pm}$ carries a representation of the corner symmetry algebra and $\mathcal{K}_{\mathcal{N}}$ describes the kinematical data at the three-dimensional null boundary $\mathcal{N}$, which is bounded by $\mathcal{C}_+$ and $\mathcal{C}_-$. The Hilbert spaces $\mathcal{K}_{\mathcal{C}_\pm}$ at the two corners are related by conjugation, i.e.\ $\mathcal{K}_{\mathcal{C}_+}$ is the dual space of $\mathcal{K}_{\mathcal{C}_-}=\mathcal{K}_{\mathcal{C}_+}^*$. A generic element of the kinematical boundary state will then admit the tensor decomposition:
\begin{align}
|\Psi\rangle\equiv\underset{AA'}{\int\mathllap{\sum}}|A\rangle\otimes|{\Psi}^{A}_{A'}\rangle\otimes\langle A'|\in \sfit{K}_{\mathcal{N}},
\end{align}
where $\{|A\rangle,|B\rangle,\dots\}$ is a basis of $\mathcal{C}_+$ and $\{\langle A|,\langle B|,\dots\}$ is the dual basis at $\mathcal{C}_-$. In the following, the boundary labels $A,B,\dots$ play the role of edge modes \cite{Balachandran:1994up,PhysRevD.51.632,DonnFreid,Donnelly:2016rvo,Wieland:2017zkf,Wieland:2017cmf,Wieland:2021vef,Gomes:2016mwl,Speranza:2017gxd,Geiller:2017xad,Takayanagi:2019tvn,Francois:2021aa,Freidel:2019ees,Freidel:2020xyx,PhysRevLett.128.171302,Freidel:2021cjp,Donnelly:2020xgu,Goeller:2022rsx,Giesel:2024xtb} for the underlying boundary symmetries.\smallskip

At this point, we can consider two cases depending on whether $\mathcal{C}_+$ lies in the causal future or past of  $\mathcal{C}_-$.
At the classical level, the two cases represent two disjoint sectors of the gravitational phase space related by parity \cite{Wieland:2025qgx}. A split of the classical phase space into disjoint regions is the classical analogue of a decomposition of the Hilbert space into orthogonal parts. We can thus expect that the kinematical Hilbert space splits into components according to \begin{equation}
 \mathcal{K}_{\mathcal{N}}=\mathcal{K}_{\mathcal{N}}^\uparrow\oplus\mathcal{K}_{\mathcal{N}}^\downarrow.\label{orthodecomp}
 \end{equation}
This expectation is confirmed by recent results on the quantisation of gravitational null initial data in terms of an auxiliary conformal field theory, in which the two sectors appear as two different inequivalent representations of the same boundary symmetry algebra, see \cite{Wieland:2025qgx,Wieland:2024dop}. To distinguish the two sectors, we use the following condensed notation
\begin{align}
|\gamma,\uparrow\rangle\equiv\underset{AA'}{\int\mathllap{\sum}}|A\rangle\otimes|\ou{\gamma}{A}{A'}\rangle\otimes\langle A'|\in \sfit{K}_{\mathcal{N}}^\uparrow,\label{state1}\\
|\gamma,\downarrow\rangle\equiv\underset{AA'}{\int\mathllap{\sum}}|A\rangle\otimes|\uo{\gamma}{A'}{A}\rangle\otimes\langle A'|\in\sfit{K}_{\mathcal{N}}^\downarrow,\label{state2}
\end{align}
in which the relative position of the indices indicates wether $\mathcal{C}_+$ lies in the causal future ($|\gamma,\uparrow$) or past $(|\gamma,\downarrow)$ of $\mathcal{C}_-$. In addition,
\begin{equation}
\sfit{K}_{\mathcal{N}}^{\uparrow,\downarrow}=\mathcal{K}_{\mathcal{C}_+}\otimes\mathcal{K}_{\mathcal{N}}^{\uparrow,\downarrow}\otimes\mathcal{K}_{\mathcal{C}_-}.
\end{equation}
In the following, we assume that there exists a discrete symmetry that can flip the orientation, thus mapping an outgoing null surface into an infalling  null surface. We denote this map as
\begin{align}
\varPi|\gamma,\uparrow\rangle=|\gamma,\downarrow\rangle,\qquad\varPi^2=\bbvar{1}.
\end{align}
To simplify our presentation, we use the following condensed tensor (DeWitt) notation. We identify the states $|\gamma,\uparrow\rangle$ and $|\gamma,\downarrow\rangle$ with their tensor-valued components $|\ou{\gamma}{A}{A'}\rangle$ and $|\uo{\gamma}{A'}{A}\rangle$ in $\mathcal{K}_\mathcal{N}$. States with opposite orientation are then identified as $\varPi|\ou{\gamma}{A}{A'}\rangle=|\uo{\gamma}{A'}{A}\rangle$. Using Einstein's summation convention, we sum or integrate\footnote{The DeWitt notation does not distinguish between finite and infinite-dimensional Hilbert spaces and between summation and integration \cite{kiefer}.} over repeated upper and lower $A, B,C,\dots$ indices. As mentioned above, the  relative position of the indices indicates whether we are considering an infalling or outgoing null surface. See \hyperref[fig2]{\color{darkblue} Figure 3} below.

\begin{figure}[h]
\centering
\includegraphics[width=0.75\textwidth]{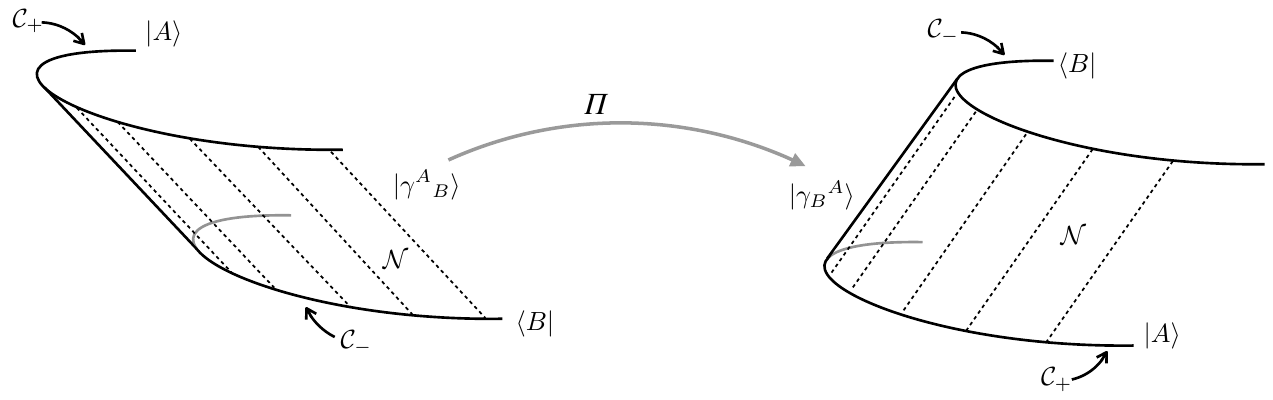}
\caption{Index conventions for boundary states distinguish the two orientation of the null slab $\mathcal{N}$. \label{fig3}}
\end{figure}%

Moving forward, let us briefly consider the Hilbert space inner product, which introduces a natural isomorphism between the Hilbert space and its dual space. 
\begin{align}
|\alpha,\beta\rangle&=\underset{AA'}{\int\mathllap{\sum}}\frac{1}{\sqrt{2}}\left(|A\rangle\otimes|\ou{\alpha}{A}{A'}\rangle\otimes\langle A'|+|A\rangle\otimes|\uo{\beta}{A'}{A}\rangle\otimes\langle A'|\right)\in\sfit{K}_{\mathcal{N}},\\
\langle\alpha,\beta|&=\underset{AA'}{\int\mathllap{\sum}}\frac{1}{\sqrt{2}}\left(\langle A|\otimes\langle\uo{\alpha}{A}{A'}|\otimes| A'\rangle+\langle A|\otimes\langle\ou{\beta}{A'}{A}|\otimes| A'\rangle\right)\in\sfit{K}_{\mathcal{N}}^\dagger.
\end{align}
The inner product between two such states which are quantum superpositions of infalling and outgoing null slabs of geometry is then given by
\begin{equation}
\langle\gamma,\delta|\alpha,\beta\rangle=\frac{1}{2}\left(\langle\uo{\gamma}{{A}}{{A}'}|\ou{\alpha}{A}{A'}\rangle +\langle\ou{\delta}{{A}'}{{A}}|\uo{\beta}{A'}{A}\rangle \right).
\end{equation}

 One of our key assumption is that the Hilbert space at the corner carries a representation of the generators $Q_\xi[\mathcal{C}]$ of the corner symmetry algebra. We can thus introduce the reduced matrix elements of the generators. In our notation
\begin{equation}
\ou{Q}{A}{B\xi}[\mathcal{C}]=\langle {A}|Q_\xi[\mathcal{C}]B\rangle.
\end{equation}
In addition, we assume that the representation of the corner symmetry algebra is unitary, $\langle {A}|Q_\xi[\mathcal{C}]B\rangle=\langle Q_\xi[\mathcal{C}]{A}|B\rangle$. 

In here, we leave the corner symmetry algebra $\mathfrak{g}_{\mathcal{C}_\pm}$ unspecified. Natural choices are the tangent bundles $T\mathcal{C}_\pm$ equipped with the natural Lie bracket bracket of vector fields and the jet bundles that describe the normal expansion of tangent vectors $\xi^a\in T\mathcal{M}\big|_{\mathcal{C}_\pm}$ with respect to the two transversal directions to the corner, see e.g.\ \cite{Neri:2025fsh}, which contains a concise review of the construction.
\subsection{Projector onto physical states}
\noindent We obtain the physical states at the null boundary by imposing the flux-balance laws \eref{charge-flx}. At the quantum level, the restriction to physical states is realized through a generalized projector,
\begin{align}
\boldsymbol{P}:&|\ou{\gamma}{A}{A'}\rangle\rightarrow|\ou{\gamma}{A}{A'}\rangle:=\ou{P}{AB'}{A'B}|\ou{\gamma}{B}{B'}\rangle,\label{P-def}\\
&|\uo{\gamma}{A'}{A}\rangle\rightarrow|\ou{\gamma}{A}{A'}\rangle:=\ou{P}{AB'}{A'B}|\uo{\gamma}{B'}{B}\rangle.\label{P-def2}
\end{align}
The conditions for this map to project onto the solution space of the constraints is
\begin{align}
\ou{Q}{A}{C{\xi}}[\mathcal{C}_+]\ou{P}{CB'}{A'B}-\ou{Q}{C'}{A'\xi}[\mathcal{C}_-]\ou{P}{AB'}{C'B}=H_\xi[\mathcal{N}]\ou{P}{AB'}{A'B},\label{Pinv1}\\
\ou{P}{AB'}{A'C}\ou{Q}{C}{B\xi}[\mathcal{C}_+]-\ou{P}{AC'}{A'A}\ou{Q}{B'}{C'\xi}[\mathcal{C}_-]=\ou{P}{AB'}{A'B}H_\xi[\mathcal{N}],\label{Pinv2}
\end{align}
for all symmetry algebra-valued test functions $\xi^a$ on $\mathcal{N}$. The matrix elements of this projector define the physical inner product. Consider  two kinematical states
\begin{equation}
\Psi=\underset{AA'}{\int\mathllap{\sum}}|A\rangle\otimes|\ou{\psi}{A}{A'}\rangle\otimes\langle A'|,\qquad 
\Phi=\underset{AA'}{\int\mathllap{\sum}}|A\rangle\otimes|\ou{\phi}{A}{A'}\rangle\otimes\langle A'|
\end{equation}
The physical inner product on a null surface is then given by
\begin{equation}
\langle\Phi|\Psi\rangle_{\mtext{phys}}=\langle\uo{\phi}{B}{B'}|\ou{{P}}{BA'}{B'A}|\ou{\psi}{A}{A'}\rangle\label{phys-prod}
\end{equation}
Since there are gauge symmetries, the physical inner product \eref{phys-prod} between the kinematical boundary states $\Phi,\Psi\in\sfit{K}_{\mathcal{N}}$ contains a kernel. Two kinematical states $\Psi$ and $\Psi'$ are said to be gauge equivalent, i.e.\ $\Psi\sim\Psi'$ if their difference is a null vector, i.e.\ $\forall\Phi\in\sfit{K}_{\mathcal{N}}:\langle\Phi|\Psi-\Psi'\rangle_{\mtext{phys}}=0$. The physical Hilbert space is then given by the completion of the space of equivalence classes $[\Psi]$ under the norm induced by the inner product.\smallskip

At the classical level, the corner charges $Q_\xi[\mathcal{C}_\pm]$ are Dirac observables, which are conjugate to edge modes for the underlying gauge symmetries \cite{Balachandran:1994up,PhysRevD.51.632,DonnFreid,Donnelly:2016rvo,Wieland:2017zkf,Wieland:2017cmf,Wieland:2021vef,Gomes:2016mwl,Speranza:2017gxd,Geiller:2017xad,Takayanagi:2019tvn,Francois:2021aa,Freidel:2019ees,Freidel:2020xyx,PhysRevLett.128.171302,Freidel:2021cjp,Donnelly:2020xgu,Goeller:2022rsx,Giesel:2024xtb}. This means that they commute with the projector onto physical states.  This is the same as to say 
\begin{equation}
\langle\Phi|Q_\xi[\mathcal{C}_\pm]\Psi\rangle_{\mtext{phys}}=\langle Q_\xi[\mathcal{C}_\pm]\Phi|\Psi\rangle_{\mtext{phys}},
\end{equation}
for all kinematical states $\Phi,\Psi\in\sfit{K}_{\mathcal{N}}$ and symmetry algebra-valued test functions $\xi^a$ on $\mathcal{N}$. In other words, the projector $\ou{P}{AB'}{BA'}$ is invariant under the edge symmetries
\begin{align}
\ou{Q}{A}{C{\xi_+}}[\mathcal{C}_+]\ou{P}{CB'}{A'B}-\ou{P}{AB'}{A'C}\ou{Q}{C}{B\xi_+}[\mathcal{C}_+]=0,\label{Psymm1}\\
\ou{P}{AC'}{A'B}\ou{Q}{B'}{C'\xi_-}[\mathcal{C}_-]-\ou{Q}{C'}{A'\xi_-}[\mathcal{C}_-]\ou{P}{AB'}{C'B}=0.\label{Psymm2}
\end{align}

Finally, we assume that there exist special bulk and boundary states. First of all, we assume that there exists an edge vacuum $T^A\in\mathcal{K}_{\mathcal{C}_+}$ and $(T^\dagger_A\in\mathcal{K}_{\mathcal{C}_-}$) such that
\begin{equation}
\ou{Q}{E}{B\xi}[\mathcal{C}_+]{T}^{B}=0\quad \text{and}\quad \ou{Q}{B}{A\xi}[\mathcal{C}]{T}^\dagger_{B}=0,  \label{corner-vac}
\end{equation}
where $T^\dagger_A$ and $T^A$ are related by the Hilbert space inner product of $\mathcal{K}_{\mathcal{C}_\pm}$.
We expect that the edge state $T^A$ represents a quantum state of the boundary algebra in which the two-dimensional corner $\mathcal{C}_\pm$ shrinks into a point (tip). In addition, we assume that there is a bulk vacuum $|\ou{\Omega}{A}{B'}\rangle$ (and $|\uo{\Omega}{B'}{A}\rangle$), which can serve as an interwiner for the corner symmetries
\begin{equation}
\forall \xi^a_+\in \mathfrak{g}_{\mathcal{C}_+}\,\exists \,\xi^a_-\in \mathfrak{g}_{\mathcal{C}_-}:\ou{Q}{A}{B\xi_+}[\mathcal{C}_+]|\ou{\Omega}{B}{A'}\rangle=\ou{Q}{B'}{A'\xi_-}[\mathcal{C}_-]|\ou{\Omega}{A}{B'}\rangle,\label{Om-int}
\end{equation}
in which the mapping between $\xi^a_+$ and $\xi^a_-$, which we assume to be linear, defines a sort of parallel transport $h[\mathcal{C}_-\rightarrow\mathcal{C}_+]:\mathfrak{g}_{\mathcal{C}_-}\rightarrow \mathfrak{g}_{\mathcal{C}_+}$, $\xi^a_+=\ou{h}{a}{a'}[\mathcal{C}_-\rightarrow\mathcal{C}_+]\xi^{a'}_-$.\footnote{It seems possible that this holonomy can take values in the algebra of operators on the Hilbert space of radiative modes $\sfit{K}_{\mathcal{N}}$. In this case, $\ou{h}{a}{a'}$ would be operator-valued, and only expectation values $\langle\psi|\ou{h}{a}{a'}|\psi\rangle$ for $\psi\in\mathit{K}_{\mathcal{N}}$ would define ordinary classical holonomies.} 
Finally, we also assume that the vacuum state is a physical state that satisfies all constraints, i.e.\
\begin{equation}
H_\xi[\mathcal{N}]|\ou{\Omega}{A}{A'}\rangle=\ou{Q}{A}{B\xi}[\mathcal{C}_+]|\ou{\Omega}{B}{A'}\rangle-\ou{Q}{B'}{A'\xi}[\mathcal{C}_-]|\ou{\Omega}{A}{B'}\rangle,\quad\forall\xi^a\in \mathfrak{g}_{\mathcal{N}}.\label{blk-vac}
\end{equation}
Notice that  no special matching condition between the values of the smearing functions $\xi^a_\pm=\xi^a\big|_{\mathcal{C}_\pm}$ need to be imposed here.

\subsection{Proposal for the amplitudes}
\begin{figure}[h]
\centering
\begin{subfigure}{0.45\textwidth}\centering
\includegraphics[height=13em]{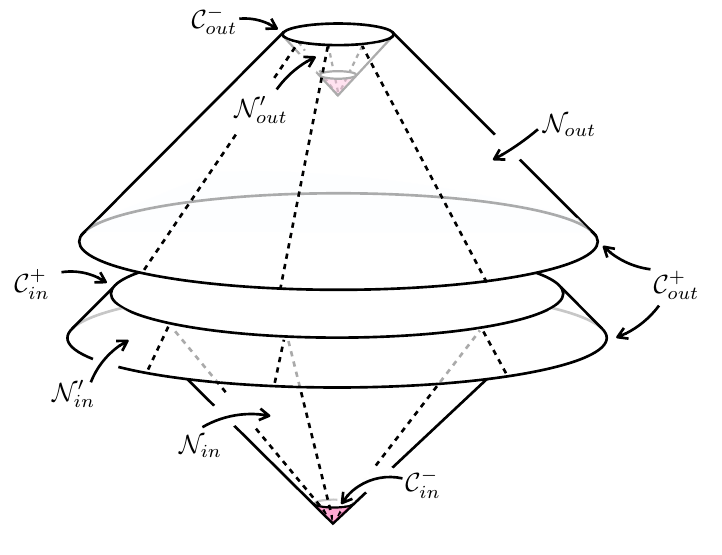}
\caption{Subdivision of a causal diamond}
\end{subfigure}
\begin{subfigure}{0.45\textwidth}\centering
\includegraphics[height=13em]{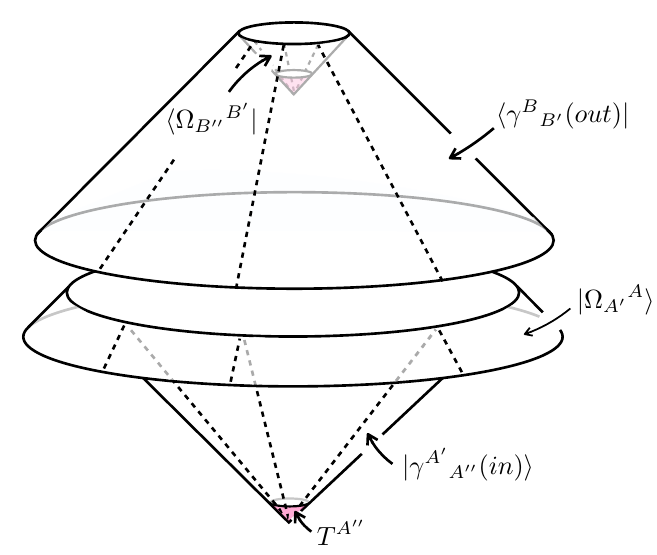}
\caption{Boundary states}
\end{subfigure}
\caption{We consider a causal diamond whose boundary is divided into past and future null hypersurfaces $\mathcal{N}_{\mtext{in}}$ and $\mathcal{N}_{\mtext{out}}$ (left panel).  To assign states and amplitudes to the causal diamond, we introduce two further auxiliary null slabs: $\mathcal{N}_{\mtext{in}}'$ is an infalling null surface, which is attached to the outgoing null surface $\mathcal{N}_{\mtext{in}}$ at its future two-dimensional boundary $\mathcal{C}^+_{\mtext{in}}$. In addition, there is $\mathcal{N}_{\mtext{out}}'$ which is an outgoing null surface attached to the top boundary of the infalling null surface $\mathcal{N}_{\mtext{out}}$. The two null surfaces $\mathcal{N}_{\mtext{out}}\cup\mathcal{N}_{\mtext{out}}'$ and $\mathcal{N}_{\mtext{in}}'\cup\mathcal{N}_{\mtext{in}}$ are diffeomorphic and define, therefore, the same abstract null surface $\mathcal{N}$. In a diffeomorphism invariant theory, the dynamics is imposed by a general projector $\boldsymbol{P}$ onto physical states. This projector maps kinematical boundary states at $\mathcal{N}\simeq\mathcal{N}_{\mtext{out}}\cup\mathcal{N}_{\mtext{in}}'\simeq\mathcal{N}_{\mtext{in}}'\cup\mathcal{N}_{\mtext{in}}$ into the same physical Hilbert space $\sfit{K}_{\mtext{phys}}^{\mathcal{N}}$. The transition amplitudes are the matrix elements of this projector between the respective \emph{in}-states and \emph{out}-states, which are given by schematically $|\mtext{in}\rangle=|\Omega,\downarrow\rangle\otimes|\gamma(\mtext{in}),\uparrow\rangle$ and $|\mtext{out}\rangle=|\gamma(\mtext{out}),\downarrow\rangle\otimes |\Omega,\uparrow\rangle$ (right panel). Amplitudes for physical processes are then merely given by the matrix elements $\boldsymbol{A}(\mtext{in}\rightarrow\mtext{out})=\langle\mtext{out}|\boldsymbol{P}|\mtext{in}\rangle$. The two vertices at the bottom and top of the causal diamond are removed from $\mathcal{N}_{\mtext{in}}$ and $\mathcal{N}_{\mtext{out}}$ and replaced by corner states $T^A\in\mathcal{K}_{\mathcal{C}_\pm}$ (shaded regions).  \label{fig4}}
\end{figure}

\noindent In this section, we present a proposal for how to build local quantum gravitational amplitudes in causal diamonds from elementary elements (states, vacua, projectors) that carry a representation of the boundary symmetry algebra. The elementary building blocks required to establish the construction are summarized below.
\begin{enumerate}
\item[(i)] A kinematical boundary Hilbert space $\sfit{K}_{\mathcal{N}}=\mathcal{K}_{\mathcal{C}_+}\otimes\mathcal{K}_{\mathcal{N}}\otimes\mathcal{K}_{\mathcal{C}_-}$ for each null slab, which is the tensor product of bulk and boundary modes. In here, the Hilbert space of the edge modes $\mathcal{K}_{\mathcal{C}_\pm}$ carries a quantum representation of the gravitational boundary charges $Q_\xi[\mathcal{C}_\pm]$ and $\mathcal{K}_{\mathcal{N}}$ contains the radiative data. The two boundary factors are related by hermitian conjugation, i.e.\ $\mathcal{K}_{\mathcal{C}_+}=\mathcal{K}_{\mathcal{C}_-}^\dagger$ and they represent the corner data at the upper (lower) boundary of $\mathcal{N}$, i.e.\ $\partial\mathcal{N}=\mathcal{C}_+\cup\mathcal{C}_-^{-1}$.
\item[(ii)] An orthogonal decomposition of $\sfit{K}_{\mathcal{N}}=\mathcal{K}_{\mathcal{C}_+}\otimes(\mathcal{K}_{\mathcal{N}}^\uparrow\oplus\mathcal{K}_{\mathcal{N}}^\downarrow)\otimes\mathcal{K}_{\mathcal{C}_-}\equiv\sfit{K}_{\mathcal{N}}^\uparrow\oplus\sfit{K}_{\mathcal{N}}^\downarrow$, in which elements of $\mathcal{K}_{\mathcal{N}}^\uparrow$ represent outgoing null surfaces and $\mathcal{K}_{\mathcal{N}}^\downarrow$ represent infalling null surfaces. To simplify the notation, we use an abstract index notation in which $\{|A\rangle,|B\rangle,\dots\}$ is a basis of $\mathcal{K}_{\mathcal{C}_+}$ and $\{\langle A|,\langle B|,\dots\}$ is the corresponding dual basis of $\mathcal{K}_{\mathcal{C}_-}$. Accordingly, we denote elements of $\sfit{K}^\uparrow$ as $|\ou{\psi}{A}{B}\rangle\in\sfit{K}^\uparrow$ and $|\uo{\psi}{B}{A}\rangle\in\sfit{K}^\downarrow$. Repeated indices are summed over.
\item[(iii)] An algebra of constraints for each null slab generated by the flux-balance laws $C_\xi=Q_\xi[\mathcal{C}_+]-Q_\xi[\mathcal{C}_-]-H_\xi[\mathcal{N}]=0$ for all gauge elements $\xi\in \mathfrak{g}_{\mathcal{N}}$, in which $\mathfrak{g}_{\mathcal{N}}$ labels the gauge symmetry algebra on $\mathcal{N}$.
\item[(iv)] A parity swap operator $\boldsymbol{\varPi}:\sfit{K}_{\mathcal{N}}^{\uparrow,\downarrow}\rightarrow\sfit{K}_{\mathcal{N}}^{\downarrow,\uparrow}$, $\boldsymbol{\varPi}^2=\bbvar{1}$ that commutes with the constraints $[C_\xi,\boldsymbol{\varPi}]=0$ for all $\xi\in\mathfrak{g}_{\mathcal{N}}$.
\item[(v)] A projector $\boldsymbol{P}$ on $\sfit{K}^\uparrow\ni|\ou{\gamma}{A}{A'}\rangle\rightarrow \boldsymbol{P}|\ou{\gamma}{A}{A'}\rangle=\ou{P}{AB'}{A'B}|\ou{\gamma}{B}{B'}\rangle$ (and equally for states in $\sfit{K}^\downarrow$) onto the solution space of the constraints, i.e.\ $\boldsymbol{P}C_\xi=C_\xi \boldsymbol{P}=0$, that commutes with the boundary charges, i.e.\ $[Q_\xi[\mathcal{C}_\pm],\boldsymbol{P}]=0$. Thus, the boundary charges are physical observables. 

\item[(vi)] A no-geometry edge vacuum $T^A\in\mathcal{K}_{\mathcal{C}_\pm}$ that is annihilated by all boundary charges, i.e.\ $\ou{Q}{A}{B\xi}[\mathcal{C}_\pm]T^B=0$  for all $ \xi\in\mathfrak{g}_{\mathcal{C}_\pm}$. The edge vacuum represents a state in which the  geometry of the boundary $\mathcal{C}$ shrinks to a point (tip).
\item[(vii)] An outgoing bulk vacuum $|\ou{\Omega}{A}{A'}\rangle\in\sfit{K}^\uparrow$ and an infalling bulk vacuum $|\uo{\Omega}{A'}{A}\rangle\in\sfit{K}^\downarrow$ which are physical states that satisfy all constraints, i.e.\  $\forall\xi\in \mathfrak{g}_{\mathcal{N}}:$ $H_\xi[\mathcal{N}]|\ou{\Omega}{A}{A'}\rangle=\ou{Q}{A}{B\xi}[\mathcal{C}_+]|\ou{\Omega}{B}{A'}\rangle-\ou{Q}{B'}{A'\xi}[\mathcal{C}_-]|\ou{\Omega}{A}{B'}\rangle=0$ and equally for $|\uo{\Omega}{A'}{A}\rangle$. These states  behave as intertwiners  for the corner symmetry group, i.e.\ the vacuum state satisfies \eref{Om-int} and \eref{blk-vac}.
\end{enumerate}
We take (i--vii) as basic axioms for a top-down approach to local amplitudes. From the bottom-up perspective, the individual elements need to be realised on appropriate bulk and corner Hilbert spaces, as introduced in e.g.\ \cite{Wieland:2025qgx}. Given these elements, we can build transition amplitudes for kinematical boundary states $|\ou{\gamma}{A}{A'}(\mtext{in},\mtext{out})\rangle$. We set
\begin{align}
&\boldsymbol{A}\big(\gamma(\mtext{in})\rightarrow\gamma(\mtext{out})\big)=\nonumber\\
&=\langle\ou{\gamma}{B'}{B}(\mtext{out})|\ou{P}{BD'}{B'D}|\uo{\Omega}{C'}{D}\rangle\,\langle \uo{\Omega}{D'}{D''}|\ou{P}{C'A''}{C''A'}|\ou{\gamma}{A'}{A''}(\mtext{in})\rangle {T}^\dagger_{D''}T^{C''},\label{ampl-def}
\end{align}
in which the pattern of index contractions follows the geometry of an abstract causal diamond. See \hyperref[fig4]{\color{darkblue}Figure 4} for an illustration. The proposal \eref{ampl-def} is motivated by the observation that the transition amplitudes for a diffeomorphism invariant theory are matrix elements of a general projector that maps kinematical boundary states into physical states on an abstract Cauchy surface \cite{rovelli, zakolec, alexreview, Rovelli:2015gwa}. To understand the pattern of index contractions, we can imagine two auxiliary null surfaces that are glued to the top (bottom) part of the past (future) boundary of the causal diamond (\hyperref[fig4]{\color{darkblue}Figure 4}).  In this way, it is possible to prescribe the \emph{in}-states and the \emph{out}-states on the same abstract (diffeomorphically equivalent) null surfaces. In the construction, the two vertices at the bottom and top of the causal diamond are replaced by finite boundaries $\mathcal{C}^-_{\mtext{out}}$ and $\mathcal{C}^+_{\mtext{out}}$, which are then glued to appropriate edge states $T^A$, which are assumed to be singlets (vacuum states) for the corner symmetry algebra. 

To understand the properties of this proposal, we consider first the action of the constraints. We obtain
\begin{align}
&\boldsymbol{A}\big(H_\xi[\mathcal{N}_{\mtext{in}}]\gamma(\mtext{in})\rightarrow\gamma(\mtext{out})\big)=\langle\ou{\gamma}{B'}{B}(\mtext{out})|\ou{P}{BD'}{B'D}|\uo{\Omega}{C'}{D}\rangle\nonumber\\
&\qquad\times\langle \uo{\Omega}{D'}{D''}|\ou{P}{C'A''}{C''A'}H_\xi[\mathcal{N}_{\mtext{in}}]|\ou{\gamma}{A'}{A''}(\mtext{in})\rangle {T}^\dagger_{D''}T^{C''}=\nonumber\\
&=\langle\ou{\gamma}{B'}{B}(\mtext{out})|\ou{P}{BD'}{B'D}|\uo{\Omega}{C'}{D}\rangle\langle \uo{\Omega}{D'}{D''}|\ou{P}{C'A''}{C''E'}\ou{Q}{E'}{A'}_{\xi}[\mathcal{C}_{\mtext{in}}^+]|\ou{\gamma}{A'}{A''}(\mtext{in})\rangle {T}^\dagger_{D''}T^{C''}\nonumber\\
&\qquad-\langle\ou{\gamma}{B'}{B}(\mtext{out})|\ou{P}{BD'}{B'D}|\uo{\Omega}{C'}{D}\rangle\nonumber\\
&\qquad\times\langle \uo{\Omega}{D'}{D''}|\ou{P}{C'E''}{C''A'}\ou{Q}{A''}{E''}_{\xi}[\mathcal{C}_{\mtext{in}}^-]|\ou{\gamma}{A'}{A''}(\mtext{in})\rangle {T}^\dagger_{D''}T^{C''}.
\end{align}
The terms in the last two lines vanish, which follows from
\begin{align}
\ou{P}{C'E''}{C''A'}\ou{Q}{A''}{E''}[\mathcal{C}_{\mtext{in}}^-] {T}^{C''}=
\ou{P}{C'A''}{E''A'}\ou{Q}{E''}{C''}[\mathcal{C}_{\mtext{in}}^-] {T}^{C''}=0,
\end{align}
which is a consequence of \eref{Psymm2} and our assumption of the existence of an edge vacuum state
\begin{equation}
\ou{Q}{E}{C\xi}[\mathcal{C}]{T}^{C}=0.\label{edge-vac}
\end{equation}
Given these assumptions, we obtain
\begin{align}
&\boldsymbol{A}\big(H_\xi[\mathcal{N}_{\mtext{in}}]\gamma(\mtext{in})\rightarrow\gamma(\mtext{out})\big)=
\boldsymbol{A}\big(Q_\xi[\mathcal{C}_{\mtext{in}}^{+}]\gamma(\mtext{in})\rightarrow\gamma(\mtext{out})\big).\label{HinA}
\end{align}
In addition, we then also have
\begin{align}
&\boldsymbol{A}\big(\gamma(\mtext{in})\rightarrow H_\xi[\mathcal{N}_{\mtext{out}}]\gamma(\mtext{out})\big)=\boldsymbol{A}\big(\gamma(\mtext{in})\rightarrow Q_\xi[\mathcal{C}^{\mtext{out}}_+]\gamma(\mtext{out})\big).\label{HoutA}
\end{align}
This can be seen by first noting that
\begin{align}
&\boldsymbol{A}\big(\gamma(\mtext{in})\rightarrow H_\xi[\mathcal{N}_{\mtext{out}}]\gamma(\mtext{out})\big)=\langle\ou{\gamma}{B'}{B}(\mtext{out})|H_\xi[\mathcal{N}_{\mtext{out}}]\ou{P}{BD'}{B'D}|\uo{\Omega}{C'}{D}\rangle\nonumber\\
&\qquad\times\langle \uo{\Omega}{D'}{D''}|\ou{P}{C'A''}{C''A'}|\ou{\gamma}{A'}{A''}(\mtext{in})\rangle {T}^\dagger_{D''}T^{C''}=\nonumber\\
&=\langle\ou{\gamma}{B'}{B}(\mtext{out})|\ou{Q}{B}{E}_{\xi}[\mathcal{C}_{\mtext{out}}^+]\ou{P}{ED'}{B'D}|\uo{\Omega}{C'}{D}\rangle\langle \uo{\Omega}{D'}{D''}|\ou{P}{C'A''}{C''A'}|\ou{\gamma}{A'}{A''}(\mtext{in})\rangle {T}^\dagger_{D''}T^{C''}\nonumber\\
&\qquad-\langle\ou{\gamma}{B'}{B}(\mtext{out})|\ou{Q}{E'}{B'}_{\xi}[\mathcal{C}_{\mtext{out}}^-]\ou{P}{BD'}{E'D}|\uo{\Omega}{C'}{D}\rangle\nonumber\\
&\qquad\times\langle \uo{\Omega}{D'}{D''}|\ou{P}{C'A''}{C''A'}|\ou{\gamma}{A'}{A''}(\mtext{in})\rangle {T}^\dagger_{D''}T^{C''}.
\end{align}
Using the intertwining properties of the projector, i.e. \eref{Om-int}, and the invariance conditions of the bulk and boundary vacuum, i.e.\ \eref{Psymm2} and \eref{edge-vac}, it is easy to see that the second part of this equation vanishes,
\begin{align}
&\langle\ou{\gamma}{B'}{B}(\mtext{out})|\ou{Q}{E'}{B'}_{\xi}[\mathcal{C}_{\mtext{out}}^-]\ou{P}{BD'}{E'D}|\uo{\Omega}{C'}{D}\rangle\langle \uo{\Omega}{D'}{D''}|\ou{P}{C'A''}{C''A'}|\ou{\gamma}{A'}{A''}(\mtext{in})\rangle {T}^\dagger_{D''}=\nonumber\\
&=\langle\ou{\gamma}{B'}{B}(\mtext{out})|\ou{P}{BE'}{B'D}|\uo{\Omega}{C'}{D}\rangle\langle \uo{\Omega}{D'}{D''}|\ou{Q}{D'}{E'}_{\xi}[\mathcal{C}_{\mtext{out}}^-]\ou{P}{C'A''}{C''A'}|\ou{\gamma}{A'}{A''}(\mtext{in})\rangle {T}^\dagger_{D''}=\nonumber\\
&=\langle\ou{\gamma}{B'}{B}(\mtext{out})|\ou{P}{BE'}{B'D}|\uo{\Omega}{C'}{D}\rangle\nonumber\\
&\qquad\times\langle \uo{\Omega}{E'}{E''}|\ou{Q}{D''}{E''}_{h[\mathcal{C}_{\mtext{in}}^+\rightarrow\mathcal{C}_{\mtext{in}}^-]\xi}[\mathcal{C}_{\mtext{in}}^-]\ou{P}{C'A''}{C''A'}|\ou{\gamma}{A'}{A''}(\mtext{in})\rangle {T}^\dagger_{D''}\nonumber\\
&=\langle\ou{\gamma}{B'}{B}(\mtext{out})|\ou{P}{BE'}{B'D}|\uo{\Omega}{C'}{D}\rangle\nonumber
\\&\qquad\times\langle \uo{\Omega}{E'}{E''}|\ou{P}{C'A''}{C''A'}|\ou{\gamma}{A'}{A''}(\mtext{in})\rangle\ou{Q}{D''}{E''}_{h[\mathcal{C}_{\mtext{in}}^+\rightarrow\mathcal{C}_{\mtext{in}}^-]\xi}[\mathcal{C}_{\mtext{in}}^-]{T}^\dagger_{D''}=0.
\end{align}
In the same way, it is easy to show that the amplitudes realise local charge conservation at the corner, where the future and past states intersect, i.e.\
\begin{equation}
\boldsymbol{A}\big(Q_\xi[\mathcal{C}_{\mtext{in}}^{+}]\gamma(\mtext{in})\rightarrow\gamma(\mtext{out})\big)=
\boldsymbol{A}\big(\gamma(\mtext{in})\rightarrow Q_{h[\mathcal{C}_{\mtext{in}}^{+}\rightarrow\mathcal{C}_{\mtext{out}}^{+}]\xi}[\mathcal{C}_{\mtext{out}}^{+}]\gamma(\mtext{out})\big).\label{charge-cons}
\end{equation}

\section{Summary and discussion}
\noindent To summarize, we introduced amplitudes \eref{ampl-def} that satisfy Ward identities (\ref{HoutA}, \ref{HinA}) and charge conservation laws \eref{charge-cons} across the two-dimensional corner $\mathcal{C}_{\mtext{out}}^+$, where the \emph{in}-states and \emph{out}-states intersect, see \hyperref[fig4]{\color{darkblue} Figure 4} for an illustration. {The charge conservation laws are reminiscent of recent developments in perturbative gravity, where the gravitational $S$-matrix satisfies a tower of soft-graviton theorems, which have been generalized formally to all orders in the perturbative $r^{-1}$ expansion in the vicinity of asymptotic future (past) null infinity \cite{Strominger:2017zoo,Campiglia:2016jdj,Freidel:2021dfs}. These conservation laws follow from gravitational Ward identities\footnote{They are obtained from the pull-back of the Wheeler DeWitt equations to future (past) null infinity $\mathcal{I}_\pm$.} and impose pointwise matching conditions between local charge densities (energy, momentum and higher charge multipoles) at the asymptotic two-sphere where the future and past components of the asymptotic null boundary intersect. It is however unclear at this point whether these symmetries are stringent enough to determine the $S$-matrix elements completely.} In here, we took a top-down perspective instead. Rather than starting from a presupposed $S$-matrix, which may not exist in a full theory of quantum gravity, we merely asked how local amplitudes for the quantum gravitational field could be constructed from the underlying symmetries alone. 
 The key building block is the generalized projector \eref{P-def} onto physical states. This projector defines an intertwiner for the gauge symmetry group at the  null boundary. If we then also assume that there are certain preferred bulk and corner vacua $|\ou{\Omega}{A}{A'}\rangle$ and $T^A$ that can act as intertwiners\footnote{That the vacuum states $|\ou{\Omega}{A}{A'}\rangle$ at $\mathcal{N}$ and $T^A$ at $\mathcal{C}_\pm$ transform covariantly under the boundary symmetries can be understood as a realisation  of the \emph{quantum equivalence principle} recently introduced \cite{Hardy:2020qg,Giacomini:2019aa}.} for the corner symmetry algebra, \eref{corner-vac}, \eref{Om-int} and \eref{blk-vac}, we saw that it is possible to construct amplitudes from simple tensor contractions of the vacuum states and generalized projectors alone. The underlying pattern of index contractions is inferred from the topological gluing between the various components of the null boundary (see \hyperref[fig4]{\color{darkblue}Figure 4}). \smallskip
 
Throughout our discussion, we were deliberately vague about how to realise the proposal in practice. The key building block is the projector \eref{P-def} that maps kinematical boundary states to physical states on an abstract null boundary. Let us now provide some background material how this formal structure could be realized in practice. First of all, we shall note that the same formal structure also appears in the spinfoam and group field theory (GFTs) approaches to quantum gravity, \cite{rovelli, Ooguri:1991ni,zakolec, alexreview, Barrett:2008wh,Freidel:1998pt,LQGvertexfinite,flppdspinfoam,Frances2014,Dittrich:2014ala,Steinhaus:2020lgb,Asante:2020qpa,Asante:2022dnj,Livine2025507} and \cite{Gielen:2011dg,Carrozza:2020akv,Oriti:2017ave,oriti}, in which the transition amplitudes are matrix elements of a generalized projector on a  kinematical boundary Hilbert space. Spinfoams realise the projector through a sum over surfaces path integral between the respective \emph{in}-states and \emph{out}-states. GFTs define the projector through an auxiliary quantum field theory on the Wheelerian superspace of a single atom of space. Recent results on the quantum geometry of the null cone suggest a slightly different strategy to realise the same goal. The idea is to define the projector through the $n$-point functions of an auxiliary conformal field theory \cite{Ciambelli:2024swv,Wieland:2025qgx}. Schematically,
\begin{equation}
\langle \uo{\beta}{B}{B'}|\ou{P}{BA'}{B'A}|\ou{\alpha}{A}{A'}\rangle=\int_{\phi_i|_{r\rightarrow 0}=\alpha_i}^{\phi_i|_{r\rightarrow \infty}=\bar{\beta}_i}\mathcal{D}\phi_i\,\E^{-\int d^2z\,\mathcal{L}_{\mathrm{CFT}}[\phi_i,\di\phi_i]}.
\end{equation}
In here, we implicitly assumed a state operator correspondence between the gravitational  boundary states $|\ou{\alpha}{A}{A'}\rangle,|\ou{\beta}{A}{A'}\rangle,\dots$ and the field configurations $\alpha_i,\beta_i,\dots$ of the CFT. A proposal for how to construct this CFT compatible with earlier developments in loop quantum gravity \cite{Rovelliarea,Haggard:2023tnj,AshtekarLewandowskiArea,FernandoBarbero:2009ai} was put forward recently in \cite{Wieland:2025qgx}. This proposal is complementary to our present investigation. While we dealt here in this paper with a top-down approach to the fundamental amplitudes, the earlier proposal is based on a more standard bottom-up canonical quantisation of the gravitational phase space on a light-like boundary.  The present framework, on the other hand, is more abstract. It does not rely on a concrete Hilbert space realisation and can serve as a blueprint for local amplitudes in various approaches.

\noindent \emph{Acknowledgments.} W.W. thanks Maïté Dupuis for hospitality at Perimeter Institute for Theoretical Physics and Hans Briegel for hospitality at the Institute for Theoretical Physics at the University of Innsbruck. This research was supported in part by Perimeter Institute for Theoretical Physics. Research at Perimeter Institute is supported by the Government of Canada through the Department of Innovation, Science and Economic Development and by the Province of Ontario through the Ministry of Colleges and Universities. In addition, support by Deutsche Forschungsgemeinschaft (DFG, German
Research Foundation) is gratefully acknowledged. This research was funded in part through a Heisenberg fellowship of Deutsche Forschungsgemeinschaft (DFG, German
Research Foundation)---543301681.

\providecommand{\href}[2]{#2}\begingroup\raggedright\endgroup


\begin{thebibliography}{100}

\bibitem{PhysRev.160.1113}
B.~S. DeWitt, ``Quantum Theory of Gravity. I. The Canonical Theory,'' {\em
  Phys. Rev.} {\bf 160} (Aug, 1967) 1113--1148.

\bibitem{ashtekar}
A.~Ashtekar, {\em {Lectures on Non-Pertubative Canonical Gravity}}.
\newblock World Scientific, 1991.

\bibitem{kiefer}
C.~Kiefer, {\em Quantum Gravity}.
\newblock Cambridge University Press, Cambridge, 2005.

\bibitem{thiemann}
C.~Thiemann, {\em Introduction to Modern Canonical Quantum General Relativity}.
\newblock Cambridge University Press, 2007.

\bibitem{rovelli}
C.~Rovelli, {\em Quantum Gravity}.
\newblock Cambridge University Press, Cambridge, 2008.

\bibitem{Dittrich:2004cb}
B.~Dittrich, ``{Partial and complete observables for Hamiltonian constrained
  systems},'' {\em Gen. Rel. Grav.} {\bf 39} (2007) 1891--1927,
\href{http://arXiv.org/abs/gr-qc/0411013}{{\tt arXiv:gr-qc/0411013}}.

\bibitem{Reisenberger:2012zq}
M.~P. Reisenberger, ``{The symplectic 2-form for gravity in terms of free null
  initial data},'' {\em Class. Quant. Grav.} {\bf 30} (2013) 155022,
  \href{http://arXiv.org/abs/1211.3880}{{\tt arXiv:1211.3880}}.

\bibitem{Fuchs:2017jyk}
A.~Fuchs and M.~P. Reisenberger, ``{Integrable structures and the quantization
  of free null initial data for gravity},'' {\em Class. Quant. Grav.} {\bf 34}
  (2017), no.~18, 185003, \href{http://arXiv.org/abs/1704.06992}{{\tt
  arXiv:1704.06992}}.

\bibitem{Reisenberger:2018xkn}
M.~P. Reisenberger, ``{The Poisson brackets of free null initial data for
  vacuum general relativity},'' {\em Class. Quant. Grav.} {\bf 35} (2018),
  no.~18, 185012,
\href{http://arXiv.org/abs/1804.10284}{{\tt arXiv:1804.10284}}.

\bibitem{Wieland:2017cmf}
W.~Wieland, ``{Fock representation of gravitational boundary modes and the
  discreteness of the area spectrum},'' {\em Ann. Henri Poincar{\'e}} {\bf 18}
  (2017) 3695--3717,
\href{http://arXiv.org/abs/1706.00479}{{\tt arXiv:1706.00479}}.

\bibitem{Wieland:2025qgx}
W.~Wieland, ``{Quantum Geometry of the Light Cone: Fock Representation and
  Spectrum of Radiated Power},'' \href{http://arXiv.org/abs/2504.10802}{{\tt
  arXiv:2504.10802}}.

\bibitem{Wieland:2017zkf}
W.~Wieland, ``{New boundary variables for classical and quantum gravity on a
  null surface},'' {\em Class. Quantum Grav.} {\bf 34} (2017) 215008,
\href{http://arXiv.org/abs/1704.07391}{{\tt arXiv:1704.07391}}.

\bibitem{AndradeeSilva:2022iic}
R.~Andrade~e Silva and T.~Jacobson, ``{Causal diamonds in (2+1)-dimensional
  quantum gravity},'' {\em Phys. Rev. D} {\bf 107} (2023), no.~2, 024033,
  \href{http://arXiv.org/abs/2203.10084}{{\tt arXiv:2203.10084}}.

\bibitem{Jacobson:2022gmo}
T.~Jacobson and M.~R. Visser, ``{Entropy of causal diamond ensembles},'' {\em
  SciPost Phys.} {\bf 15} (2023), no.~1, 023,
  \href{http://arXiv.org/abs/2212.10608}{{\tt arXiv:2212.10608}}.

\bibitem{Bub:2024nan}
M.~W. Bub, T.~He, P.~Mitra, Y.~Zhang, and K.~M. Zurek, ``{Quantum Mechanics of
  a Spherically Symmetric Causal Diamond in Minkowski Spacetime},'' {\em Phys.
  Rev. Lett.} {\bf 134} (2025), no.~12, 121501,
  \href{http://arXiv.org/abs/2408.11094}{{\tt arXiv:2408.11094}}.

\bibitem{Freidel:2023bnj}
L.~Freidel, M.~Geiller, and W.~Wieland, ``{Corner symmetry and quantum
  geometry},'' in {\em Handbook of Quantum Gravity}, L.~M. Cosimo~Bambi and
  I.~Shapiro, eds.
\newblock Springer, 2023.
\newblock \href{http://arXiv.org/abs/2302.12799}{{\tt arXiv:2302.12799}}.

\bibitem{Wieland:2024dop}
W.~Wieland, ``{Quantum geometry of the null cone},''
  \href{http://arXiv.org/abs/2401.17491}{{\tt arXiv:2401.17491}}.

\bibitem{Wieland:2025LP}
W.~Wieland, ``Evidence for Planck luminosity bound in quantum gravity,'' {\em
  {Class. Quant. Grav.}} {\bf 42} (2025), no.~6, 06LT01.

\bibitem{Ciambelli:2024swv}
L.~Ciambelli, L.~Freidel, and R.~G. Leigh, ``{Quantum null geometry and
  gravity},'' {\em JHEP} {\bf 12} (2024) 028,
  \href{http://arXiv.org/abs/2407.11132}{{\tt arXiv:2407.11132}}.

\bibitem{Wieland:2020gno}
W.~Wieland, ``{Null infinity as an open Hamiltonian system},'' {\em JHEP} {\bf
  04} (2021) 095, \href{http://arXiv.org/abs/2012.01889}{{\tt
  arXiv:2012.01889}}.

\bibitem{Fiorucci:2025twa}
A.~Fiorucci, S.~Pekar, P.~Marios~Petropoulos, and M.~Vilatte,
  ``{Carrollian-holographic Derivation of BMS Flux-balance Laws},''
  \href{http://arXiv.org/abs/2505.00077}{{\tt arXiv:2505.00077}}.

\bibitem{Donnay:2022aba}
L.~Donnay, A.~Fiorucci, Y.~Herfray, and R.~Ruzziconi, ``{Carrollian Perspective
  on Celestial Holography},'' {\em Phys. Rev. Lett.} {\bf 129} (2022), no.~7,
  071602, \href{http://arXiv.org/abs/2202.04702}{{\tt arXiv:2202.04702}}.

\bibitem{Klinger:2025hjp}
M.~S. Klinger and R.~G. Leigh, ``{The Problem of Time and its Quantum
  Resolution},'' \href{http://arXiv.org/abs/2504.00152}{{\tt
  arXiv:2504.00152}}.

\bibitem{Wang:2008jy}
M.-T. Wang and S.-T. Yau, ``{Quasilocal mass in general relativity},'' {\em
  Phys. Rev. Lett.} {\bf 102} (2009) 021101,
  \href{http://arXiv.org/abs/0804.1174}{{\tt arXiv:0804.1174}}.

\bibitem{Szabados:2004vb}
L.~B. Szabados, ``{Quasi-Local Energy-Momentum and Angular Momentum in GR: A
  Review Article},'' {\em Living Rev. Rel.} {\bf 7} (2004)
4.

\bibitem{Donnelly:2016auv}
W.~Donnelly and L.~Freidel, ``{Local subsystems in gauge theory and gravity},''
  {\em JHEP} {\bf 09} (2016) 102, \href{http://arXiv.org/abs/1601.04744}{{\tt
  arXiv:1601.04744}}.

\bibitem{Giddings:2019hjc}
S.~B. Giddings, ``{Gravitational dressing, soft charges, and perturbative
  gravitational splitting},'' {\em Phys. Rev. D} {\bf 100} (2019), no.~12,
  126001, \href{http://arXiv.org/abs/1903.06160}{{\tt arXiv:1903.06160}}.

\bibitem{Carrozza:2022xut}
S.~Carrozza, S.~Eccles, and P.~A. Hoehn, ``{Edge modes as dynamical frames:
  charges from post-selection in generally covariant theories},''
  \href{http://arXiv.org/abs/2205.00913}{{\tt arXiv:2205.00913}}.

\bibitem{DeVuyst:2024fxc}
J.~De~Vuyst, S.~Eccles, P.~A. Hoehn, and J.~Kirklin, ``{Crossed products and
  quantum reference frames: on the observer-dependence of gravitational
  entropy},'' {\em JHEP} {\bf 07} (2025) 063,
  \href{http://arXiv.org/abs/2412.15502}{{\tt arXiv:2412.15502}}.

\bibitem{Kabel:2022efn}
V.~Kabel and W.~Wieland, ``{Metriplectic geometry for gravitational
  subsystems},'' {\em Phys. Rev. D} {\bf 106} (2022), no.~6, 064053,
  \href{http://arXiv.org/abs/2206.00029}{{\tt arXiv:2206.00029}}.

\bibitem{Hartle:1983ai}
J.~Hartle and S.~Hawking, ``{Wave Function of the Universe},'' {\em Phys.Rev.}
  {\bf D28} (1983)
2960--2975.

\bibitem{Griffiths:1984aa}
R.~B. Griffiths, ``Consistent histories and the interpretation of quantum
  mechanics,'' {\em Journal of Statistical Physics} {\bf 36} (1984), no.~1,
  219--272.

\bibitem{Schlick1948-SCHGKU-2}
M.~Schlick, {\em Gesetz, Kausalit\"{a}t und Wahrscheinlichkeit}.
\newblock Gerold \& Co, Vienna, 1948.

\bibitem{ChoquetBruhat:2010ih}
Y.~Choquet-Bruhat, P.~T. Chrusciel, and J.~M. Martin-Garcia, ``{The Cauchy
  problem on a characteristic cone for the Einstein equations in arbitrary
  dimensions},'' {\em Annales Henri Poincare} {\bf 12} (2011) 419--482,
\href{http://arXiv.org/abs/1006.4467}{{\tt arXiv:1006.4467}}.

\bibitem{Chrusciel:2012ap}
P.~T. Chrusciel and T.-T. Paetz, ``{The Many ways of the characteristic Cauchy
  problem},'' {\em Class. Quant. Grav.} {\bf 29} (2012) 145006,
\href{http://arXiv.org/abs/1203.4534}{{\tt arXiv:1203.4534}}.

\bibitem{Chrusciel:2022ozm}
P.~T. Chru{\'s}ciel and W.~Cong, ``{Characteristic gluing with $\Lambda$: 1.
  Linearised Einstein equations on four-dimensional spacetimes},'' {\em Beijing
  J. Pure Appl. Math.} {\bf 1} (2024), no.~2, 689--796,
  \href{http://arXiv.org/abs/2212.10052}{{\tt arXiv:2212.10052}}.

\bibitem{Aretakis:2021lzi}
S.~Aretakis, S.~Czimek, and I.~Rodnianski, ``{The characteristic gluing problem
  for the Einstein equations and applications},'' {\em Duke Math. J.} {\bf 174}
  (2025), no.~2, 355--402, \href{http://arXiv.org/abs/2107.02441}{{\tt
  arXiv:2107.02441}}.

\bibitem{Krasnov_2020}
K.~Krasnov, {\em Formulations of General Relativity: Gravity, Spinors and
  Differential Forms}.
\newblock Cambridge Monographs on Mathematical Physics. Cambridge University
  Press, 2020.

\bibitem{HennauxTeitelboim_book}
M.~Henneaux and C.~Teitelboim, {\em Quantization of Gauge Systems}.
\newblock Princeton University Press, Princeton, 1992.

\bibitem{PhysRevD.91.064043}
S.~Alexandrov and S.~Speziale, ``First order gravity on the light front,'' {\em
  Phys. Rev. D} {\bf 91} (Mar, 2015) 064043.

\bibitem{Bojowald:2016hgh}
M.~Bojowald, S.~Brahma, U.~Buyukcam, and F.~D'Ambrosio,
  ``{Hypersurface-deformation algebroids and effective spacetime models},''
  {\em Phys. Rev. D} {\bf 94} (2016), no.~10, 104032,
  \href{http://arXiv.org/abs/1610.08355}{{\tt arXiv:1610.08355}}.

\bibitem{Teitelboim1973542}
C.~Teitelboim, ``How commutators of constraints reflect the spacetime
  structure,'' {\em Annals of Physics} {\bf 79} (1973), no.~2, 542--557.

\bibitem{adm}
R.~Arnowitt, S.~Deser, and C.~Misner, {\em {The dynamics of general
  relativity}}, ch.~7, pp.~227--264.
\newblock Wiley, New York, 1962.
\newblock \href{http://arXiv.org/abs/gr-qc/0405109v1}{{\tt
  arXiv:gr-qc/0405109v1}}.

\bibitem{Hojman197688}
S.~A. Hojman, K.~Kucha{\v r}, and C.~Teitelboim, ``Geometrodynamics regained,''
  {\em Annals of Physics} {\bf 96} (1976), no.~1, 88--135.

\bibitem{Isenberg:2013iva}
J.~Isenberg, {\em {The Initial Value Problem in General Relativity}},
  pp.~303--321.
\newblock 2014.
\newblock \href{http://arXiv.org/abs/1304.1960}{{\tt arXiv:1304.1960}}.

\bibitem{Wieland:2021vef}
W.~Wieland, ``{Gravitational SL(2, \ensuremath{\mathbb{R}}) algebra on the
  light cone},'' {\em JHEP} {\bf 07} (2021) 057,
  \href{http://arXiv.org/abs/2104.05803}{{\tt arXiv:2104.05803}}.

\bibitem{Ciambelli:2023mir}
L.~Ciambelli, L.~Freidel, and R.~G. Leigh, ``{Null Raychaudhuri: Canonical
  Structure and the Dressing Time},''
  \href{http://arXiv.org/abs/2309.03932}{{\tt arXiv:2309.03932}}.

\bibitem{Rovelliarea}
C.~Rovelli and L.~Smolin, ``Discreteness of area and volume in quantum
  gravity,'' {\em Nuclear Physics B} {\bf 442} (1995), no.~3, 593--619,
  \href{http://arXiv.org/abs/gr-qc/9411005}{{\tt arXiv:gr-qc/9411005}}.

\bibitem{Haggard:2023tnj}
H.~M. Haggard, J.~Lewandowski, and H.~Sahlmann, ``{Emergence of Riemannian
  Quantum Geometry},'' in {\em Handbook of Quantum Gravity}, C.~Bambi,
  L.~Modesto, and I.~Shapiro, eds.
\newblock Springer, 2023.
\newblock \href{http://arXiv.org/abs/2302.02840}{{\tt arXiv:2302.02840}}.

\bibitem{AshtekarLewandowskiArea}
A.~Ashtekar and J.~Lewandowski, ``{Quantum theory of geometry I.: Area
  operators},'' {\em Class. Quant. Grav.} {\bf 14} (1997) A55--A82,
\href{http://arXiv.org/abs/gr-qc/9602046}{{\tt arXiv:gr-qc/9602046}}.

\bibitem{FernandoBarbero:2009ai}
G.~J. Fernando~Barbero, J.~Lewandowski, and E.~J.~S. Villasenor, ``{Flux-area
  operator and black hole entropy},'' {\em Phys. Rev. D} {\bf 80} (2009)
  044016, \href{http://arXiv.org/abs/0905.3465}{{\tt arXiv:0905.3465}}.

\bibitem{PhysRevD.27.2885}
D.~N. Page and W.~K. Wootters, ``Evolution without evolution: Dynamics
  described by stationary observables,'' {\em Phys. Rev. D} {\bf 27} (Jun,
  1983) 2885--2892.

\bibitem{PhysRevD.42.2638}
C.~Rovelli, ``Quantum mechanics without time: A model,'' {\em Phys. Rev. D}
  {\bf 42} (Oct, 1990) 2638--2646.

\bibitem{Rovelli:2009ee}
C.~Rovelli, ``{Forget time},'' {\em Found. Phys.} {\bf 41} (2011) 1475--1490,
\href{http://arXiv.org/abs/0903.3832}{{\tt arXiv:0903.3832}}.

\bibitem{Atiyah:1989vu}
M.~Atiyah, ``{Topological quantum field theories},'' {\em Inst. Hautes Etudes
  Sci. Publ. Math.} {\bf 68} (1989) 175--186.

\bibitem{Witten:1988hc}
E.~Witten, ``{(2+1)-Dimensional Gravity as an Exactly Soluble System},'' {\em
  Nucl. Phys. B} {\bf 311} (1988)
46.

\bibitem{Carlip:1998uc}
S.~Carlip, {\em {Quantum gravity in 2+1 dimensions}}.
\newblock Cambridge University Press,
2003.
\newblock

\bibitem{Barrett:1995mg}
J.~W. Barrett, ``{Quantum gravity as topological quantum field theory},'' {\em
  J. Math. Phys.} {\bf 36} (1995) 6161--6179,
  \href{http://arXiv.org/abs/gr-qc/9506070}{{\tt arXiv:gr-qc/9506070}}.

\bibitem{Oeckl2003}
R.~Oeckl, ``{A 'General boundary' formulation for quantum mechanics and quantum
  gravity},'' {\em Phys. Lett. B} {\bf 575} (2003) 318--324,
\href{http://arXiv.org/abs/hep-th/0306025}{{\tt arXiv:hep-th/0306025}}.

\bibitem{Oeckl:2005bv}
R.~Oeckl, ``{General boundary quantum field theory: Foundations and probability
  interpretation},'' {\em Adv. Theor. Math. Phys.} {\bf 12} (2008), no.~2,
  319--352,
\href{http://arXiv.org/abs/hep-th/0509122}{{\tt arXiv:hep-th/0509122}}.

\bibitem{Oeckl:2012ni}
R.~Oeckl, ``{A positive formalism for quantum theory in the general boundary
  formulation},'' {\em Found. Phys.} {\bf 43} (2013) 1206--1232,
  \href{http://arXiv.org/abs/1212.5571}{{\tt arXiv:1212.5571}}.

\bibitem{Oeckl:2016tlj}
R.~Oeckl, ``{A local and operational framework for the foundations of
  physics},'' {\em Adv. Theor. Math. Phys.} {\bf 23} (2019), no.~2, 437--592,
  \href{http://arXiv.org/abs/1610.09052}{{\tt arXiv:1610.09052}}.

\bibitem{Kiefer:1991xy}
C.~Kiefer, ``{Functional Schrodinger equation for scalar QED},'' {\em Phys.
  Rev. D} {\bf 45} (1992) 2044--2056.

\bibitem{Oeckl:2012qb}
R.~Oeckl, ``{Schr{\"o}dinger-Feynman quantization and composition of
  observables in general boundary quantum field theory},'' {\em Adv. Theor.
  Math. Phys.} {\bf 19} (2015) 451--506,
  \href{http://arXiv.org/abs/1201.1877}{{\tt arXiv:1201.1877}}.

\bibitem{Hohn:2014uvt}
P.~A. H{\"o}hn, ``{Quantization of systems with temporally varying
  discretization I: Evolving Hilbert spaces},'' {\em J. Math. Phys.} {\bf 55}
  (2014) 083508, \href{http://arXiv.org/abs/1401.6062}{{\tt arXiv:1401.6062}}.

\bibitem{Perez:2012wv}
A.~Perez, ``{The Spin Foam Approach to Quantum Gravity},'' {\em Living Rev.
  Rel.} {\bf 16} (2013) 3, \href{http://arXiv.org/abs/1205.2019}{{\tt
  arXiv:1205.2019}}.

\bibitem{Oreshkov:2012aa}
O.~Oreshkov, F.~Costa, and {\v C}.~Brukner, ``Quantum correlations with no
  causal order,'' {\em Nature Communications} {\bf 3} (2012), no.~1, 1092.

\bibitem{Araujo:2015qya}
M.~Ara\'ujo, C.~Branciard, F.~Costa, A.~Feix, C.~Giarmatzi, and v.~Brukner,
  ``{Witnessing causal nonseparability},'' {\em New J. Phys.} {\bf 17} (2015),
  no.~10, 102001, \href{http://arXiv.org/abs/1506.03776}{{\tt
  arXiv:1506.03776}}.

\bibitem{HardyOTQT:2012}
L.~Hardy, ``The operator tensor formulation of quantum theory,'' {\em
  Philosophical Transactions of the Royal Society A: Mathematical, Physical and
  Engineering Sciences} {\bf 370} (2012), no.~1971, 3385--3417,
  \href{http://arXiv.org/abs/1201.4390}{{\tt arXiv:1201.4390}}.

\bibitem{Ashtekar:1990gc}
A.~Ashtekar, L.~Bombelli, and O.~Reula, ``{The Covariant Phase Space Of
  Asymptotically Flat Gravitational Fields},'' in {\em Mechanics, Analysis and
  Geometry: 200 Years after Lagrange}, M.~Francaviglia and D.~Holm, eds.
\newblock North Holland, Amsterdam,
1990.
\newblock

\bibitem{Wald:1999wa}
R.~M. Wald and A.~Zoupas, ``{A General definition of `conserved quantities' in
  general relativity and other theories of gravity},'' {\em Phys. Rev. D} {\bf
  61} (2000) 084027,
\href{http://arXiv.org/abs/gr-qc/9911095}{{\tt arXiv:gr-qc/9911095}}.

\bibitem{Harlow:2020aa}
D.~Harlow and J.-q. Wu, ``Covariant phase space with boundaries,'' {\em Journal
  of High Energy Physics} {\bf 2020} (2020), no.~10, 146.

\bibitem{Oliveri:2019gvm}
R.~Oliveri and S.~Speziale, ``{Boundary effects in General Relativity with
  tetrad variables},''
\href{http://arXiv.org/abs/1912.01016}{{\tt arXiv:1912.01016}}.

\bibitem{Rezende:2009sv}
D.~J. Rezende and A.~Perez, ``{4d Lorentzian Holst action with topological
  terms},'' {\em Phys. Rev. D} {\bf 79} (2009) 064026,
  \href{http://arXiv.org/abs/0902.3416}{{\tt arXiv:0902.3416}}.

\bibitem{Balachandran:1994up}
A.~P. Balachandran, L.~Chandar, and A.~Momen, ``{Edge states in gravity and
  black hole physics},'' {\em Nucl. Phys. B} {\bf 461} (1996) 581--596,
\href{http://arXiv.org/abs/gr-qc/9412019}{{\tt arXiv:gr-qc/9412019}}.

\bibitem{PhysRevD.51.632}
S.~Carlip, ``Statistical mechanics of the (2+1)-dimensional black hole,'' {\em
  Phys. Rev. D} {\bf 51} (Jan, 1995) 632--637.

\bibitem{DonnFreid}
W.~Donnelly and L.~Freidel, ``{Local subsystems in gauge theory and gravity},''
\href{http://arXiv.org/abs/1601.04744}{{\tt arXiv:1601.04744}}.

\bibitem{Donnelly:2016rvo}
W.~Donnelly and S.~B. Giddings, ``{Observables, gravitational dressing, and
  obstructions to locality and subsystems},'' {\em Phys. Rev. D} {\bf 94}
  (2016), no.~10, 104038,
\href{http://arXiv.org/abs/1607.01025}{{\tt arXiv:1607.01025}}.

\bibitem{Gomes:2016mwl}
H.~Gomes and A.~Riello, ``{The observer's ghost: notes on a field space
  connection},'' {\em JHEP} {\bf 05} (2017) 017,
\href{http://arXiv.org/abs/1608.08226}{{\tt arXiv:1608.08226}}.

\bibitem{Speranza:2017gxd}
A.~J. Speranza, ``{Local phase space and edge modes for
  diffeomorphism-invariant theories},'' {\em JHEP} {\bf 02} (2018) 021,
  \href{http://arXiv.org/abs/1706.05061}{{\tt arXiv:1706.05061}}.

\bibitem{Geiller:2017xad}
M.~Geiller, ``{Edge modes and corner ambiguities in 3d Chern-Simons theory and
  gravity},'' {\em Nucl. Phys. B} {\bf 924} (2017) 312--365,
\href{http://arXiv.org/abs/1703.04748}{{\tt arXiv:1703.04748}}.

\bibitem{Takayanagi:2019tvn}
T.~Takayanagi and K.~Tamaoka, ``{Gravity Edges Modes and Hayward Term},'' {\em
  JHEP} {\bf 02} (2020) 167, \href{http://arXiv.org/abs/1912.01636}{{\tt
  arXiv:1912.01636}}.

\bibitem{Francois:2021aa}
J.~Fran{\c c}ois, ``Bundle geometry of the connection space, covariant
  Hamiltonian formalism, the problem of boundaries in gauge theories, and the
  dressing field method,'' {\em Journal of High Energy Physics} {\bf 2021}
  (2021), no.~3, 225, \href{http://arXiv.org/abs/2010.01597}{{\tt
  arXiv:2010.01597}}.

\bibitem{Freidel:2019ees}
L.~Freidel, E.~R. Livine, and D.~Pranzetti, ``{Gravitational edge modes: from
  Kac--Moody charges to Poincar{\'e} networks},'' {\em Class. Quant. Grav.}
  {\bf 36} (2019), no.~19, 195014,
\href{http://arXiv.org/abs/1906.07876}{{\tt arXiv:1906.07876}}.

\bibitem{Freidel:2020xyx}
L.~Freidel, M.~Geiller, and D.~Pranzetti, ``{Edge modes of gravity. Part I.
  Corner potentials and charges},'' {\em JHEP} {\bf 11} (2020) 026,
  \href{http://arXiv.org/abs/2006.12527}{{\tt arXiv:2006.12527}}.

\bibitem{PhysRevLett.128.171302}
L.~Ciambelli, R.~G. Leigh, and P.-C. Pai, ``Embeddings and Integrable Charges
  for Extended Corner Symmetry,'' {\em Phys. Rev. Lett.} {\bf 128} (Apr, 2022)
  171302, \href{http://arXiv.org/abs/2111.13181}{{\tt arXiv:2111.13181}}.

\bibitem{Freidel:2021cjp}
L.~Freidel, R.~Oliveri, D.~Pranzetti, and S.~Speziale, ``{Extended corner
  symmetry, charge bracket and Einstein\textquoteright{}s equations},'' {\em
  JHEP} {\bf 09} (2021) 083, \href{http://arXiv.org/abs/2104.12881}{{\tt
  arXiv:2104.12881}}.

\bibitem{Donnelly:2020xgu}
W.~Donnelly, L.~Freidel, S.~F. Moosavian, and A.~J. Speranza, ``{Gravitational
  edge modes, coadjoint orbits, and hydrodynamics},'' {\em JHEP} {\bf 09}
  (2021) 008, \href{http://arXiv.org/abs/2012.10367}{{\tt arXiv:2012.10367}}.

\bibitem{Goeller:2022rsx}
C.~Goeller, P.~A. Hoehn, and J.~Kirklin, ``{Diffeomorphism-invariant
  observables and dynamical frames in gravity: reconciling bulk locality with
  general covariance},'' \href{http://arXiv.org/abs/2206.01193}{{\tt
  arXiv:2206.01193}}.

\bibitem{Giesel:2024xtb}
K.~Giesel, V.~Kabel, and W.~Wieland, ``{Linking Edge Modes and Geometrical
  Clocks in Linearized Gravity},'' \href{http://arXiv.org/abs/2410.17339}{{\tt
  arXiv:2410.17339}}.

\bibitem{Loveridge2018}
L.~Loveridge, T.~Miyadera, and P.~Busch, ``Symmetry, Reference Frames, and
  Relational Quantities in Quantum Mechanics,'' {\em Foundations of Physics}
  {\bf 48} (2018), no.~2, 135--198.

\bibitem{Giacomini:2017zju}
F.~Giacomini, E.~Castro-Ruiz, and {\v C}.~Brukner, ``{Quantum mechanics and the
  covariance of physical laws in quantum reference frames},'' {\em Nature
  Commun.} {\bf 10} (2019), no.~1, 494,
  \href{http://arXiv.org/abs/1712.07207}{{\tt arXiv:1712.07207}}.

\bibitem{Giacomini:2019fvi}
F.~Giacomini, E.~Castro-Ruiz, and {\v C}.~Brukner, ``{Relativistic Quantum
  Reference Frames: The Operational Meaning of Spin},'' {\em Phys. Rev. Lett.}
  {\bf 123} (2019), no.~9, 090404, \href{http://arXiv.org/abs/1811.08228}{{\tt
  arXiv:1811.08228}}.

\bibitem{Vanrietvelde:2018pgb}
A.~Vanrietvelde, P.~A. Hoehn, F.~Giacomini, and E.~Castro-Ruiz, ``{A change of
  perspective: switching quantum reference frames via a perspective-neutral
  framework},'' {\em Quantum} {\bf 4} (2020) 225,
  \href{http://arXiv.org/abs/1809.00556}{{\tt arXiv:1809.00556}}.

\bibitem{Hoehn:2019owq}
P.~A. H\"ohn, A.~R. Smith, and M.~P. Lock, ``{The Trinity of Relational Quantum
  Dynamics},'' \href{http://arXiv.org/abs/1912.00033}{{\tt arXiv:1912.00033}}.

\bibitem{Castro-Ruiz:2021vnq}
E.~Castro-Ruiz and O.~Oreshkov, ``{Relative subsystems and quantum reference
  frame transformations},'' \href{http://arXiv.org/abs/2110.13199}{{\tt
  arXiv:2110.13199}}.

\bibitem{Cresto:2024fhd}
N.~Cresto and L.~Freidel, ``{Asymptotic higher spin symmetries I: covariant
  wedge algebra in gravity},'' {\em Lett. Math. Phys.} {\bf 115} (2025), no.~2,
  39, \href{http://arXiv.org/abs/2409.12178}{{\tt arXiv:2409.12178}}.

\bibitem{Cresto:2024mne}
N.~Cresto and L.~Freidel, ``{Asymptotic Higher Spin Symmetries II: Noether
  Realization in Gravity},'' \href{http://arXiv.org/abs/2410.15219}{{\tt
  arXiv:2410.15219}}.

\bibitem{Donnay:2024qwq}
L.~Donnay, L.~Freidel, and Y.~Herfray, ``{Carrollian $\mathscr Lw_{1+\infty}$
  representation from twistor space},'' {\em SciPost Phys.} {\bf 17} (2024),
  no.~4, 118, \href{http://arXiv.org/abs/2402.00688}{{\tt arXiv:2402.00688}}.

\bibitem{Neri:2025fsh}
G.~Neri and L.~Varrin, ``{Orbit method for Quantum Corner Symmetries},''
  \href{http://arXiv.org/abs/2507.10683}{{\tt arXiv:2507.10683}}.

\bibitem{zakolec}
C.~Rovelli, ``{Zakopane lectures on loop gravity},'' {\em PoS} {\bf QGQGS2011}
  (2011) 003,
\href{http://arXiv.org/abs/1102.3660}{{\tt arXiv:1102.3660}}.

\bibitem{alexreview}
A.~Perez, ``{The Spin-Foam Approach to Quantum Gravity},'' {\em Living Rev.
  Rel.} {\bf 16} (2013), no.~3,
\href{http://arXiv.org/abs/1205.2019}{{\tt arXiv:1205.2019}}.

\bibitem{Rovelli:2015gwa}
C.~Rovelli, ``{The strange equation of quantum gravity},'' {\em Class. Quant.
  Grav.} {\bf 32} (2015), no.~12, 124005,
  \href{http://arXiv.org/abs/1506.00927}{{\tt arXiv:1506.00927}}.

\bibitem{Strominger:2017zoo}
A.~Strominger, {\em {Lectures on the Infrared Structure of Gravity and Gauge
  Theory}}.
\newblock Princeton University Press, Princeton, 2018.
\newblock
\href{http://arXiv.org/abs/1703.05448}{{\tt arXiv:1703.05448}}.
\newblock

\bibitem{Campiglia:2016jdj}
M.~Campiglia and A.~Laddha, ``{Sub-subleading soft gravitons: New symmetries of
  quantum gravity?},'' {\em Phys. Lett. B} {\bf 764} (2017) 218--221,
  \href{http://arXiv.org/abs/1605.09094}{{\tt arXiv:1605.09094}}.

\bibitem{Freidel:2021dfs}
L.~Freidel, D.~Pranzetti, and A.-M. Raclariu, ``{Sub-subleading soft graviton
  theorem from asymptotic Einstein{\textquoteright}s equations},'' {\em JHEP}
  {\bf 05} (2022) 186, \href{http://arXiv.org/abs/2111.15607}{{\tt
  arXiv:2111.15607}}.

\bibitem{Hardy:2020qg}
L.~Hardy, ``Implementation of the Quantum Equivalence Principle,'' in {\em
  Progress and Visions in Quantum Theory in View of Gravity}, F.~Finster,
  D.~Giulini, J.~Kleiner, and J.~Tolksdorf, eds., pp.~189--220.
\newblock Springer International Publishing, Cham, 2020.

\bibitem{Giacomini:2019aa}
F.~Giacomini, E.~Castro-Ruiz, and {\v C}.~Brukner, ``Quantum mechanics and the
  covariance of physical laws in quantum reference frames,'' {\em Nature
  Communications} {\bf 10} (2019), no.~1, 494.

\bibitem{Ooguri:1991ni}
H.~Ooguri, ``{Partition functions and topology changing amplitudes in the 3-D
  lattice gravity of Ponzano and Regge},'' {\em Nucl. Phys. B} {\bf 382} (1992)
  276--304,
\href{http://arXiv.org/abs/hep-th/9112072}{{\tt arXiv:hep-th/9112072}}.

\bibitem{Barrett:2008wh}
J.~W. Barrett and I.~Naish-Guzman, ``{The Ponzano-Regge model},'' {\em Class.
  Quant. Grav.} {\bf 26} (2009) 155014,
\href{http://arXiv.org/abs/0803.3319}{{\tt arXiv:0803.3319}}.

\bibitem{Freidel:1998pt}
L.~Freidel and K.~Krasnov, ``{Spin foam models and the classical action
  principle},'' {\em Adv. Theor. Math. Phys.} {\bf 2} (1999) 1183--1247,
  \href{http://arXiv.org/abs/hep-th/9807092}{{\tt arXiv:hep-th/9807092}}.

\bibitem{LQGvertexfinite}
J.~Engle, E.~Livine, and C.~Rovelli, ``{LQG vertex with finite Immirzi
  parameter},'' {\em Nucl. Phys. B} {\bf 799} (2008) 136--149,
  \href{http://arXiv.org/abs/0711.0146}{{\tt arXiv:0711.0146}}.

\bibitem{flppdspinfoam}
J.~Engle, R.~Pereira, and C.~Rovelli, ``Flipped spinfoam vertex and loop
  gravity,'' {\em Nucl. Phys. B} {\bf 798} (2008) 251--290,
  \href{http://arXiv.org/abs/0708.1236v1}{{\tt arXiv:0708.1236v1}}.

\bibitem{Frances2014}
C.~Rovelli and F.~Vidotto, {\em {Covariant Loop Quantum Gravity}: {An
  Elementary Introduction to Quantum Gravity and Spinfoam Theory}}.
\newblock Cambridge Monographs on Mathematical Physics. Cambridge University
  Press, 11, 2014.

\bibitem{Dittrich:2014ala}
B.~Dittrich, ``{The continuum limit of loop quantum gravity - a framework for
  solving the theory},'' in {\em Loop Quantum Gravity, The First Thirty Years},
  A.~Abhay and J.~Pullin, eds., vol.~4.
\newblock World Scientific, 2017.
\newblock
\href{http://arXiv.org/abs/1409.1450}{{\tt arXiv:1409.1450}}.
\newblock

\bibitem{Steinhaus:2020lgb}
S.~Steinhaus, ``{Coarse Graining Spin Foam Quantum Gravity\textemdash{}A
  Review},'' {\em Front. in Phys.} {\bf 8} (2020) 295,
  \href{http://arXiv.org/abs/2007.01315}{{\tt arXiv:2007.01315}}.

\bibitem{Asante:2020qpa}
S.~K. Asante, B.~Dittrich, and H.~M. Haggard, ``{Effective Spin Foam Models for
  Four-Dimensional Quantum Gravity},'' {\em Phys. Rev. Lett.} {\bf 125} (2020),
  no.~23, 231301, \href{http://arXiv.org/abs/2004.07013}{{\tt
  arXiv:2004.07013}}.

\bibitem{Asante:2022dnj}
S.~K. Asante, B.~Dittrich, and S.~Steinhaus, ``{Spin foams, Refinement limit
  and Renormalization},'' \href{http://arXiv.org/abs/2211.09578}{{\tt
  arXiv:2211.09578}}.

\bibitem{Livine2025507}
E.~R. Livine, ``Spinfoam Models for Quantum Gravity,'' in {\em Encyclopedia of
  Mathematical Physics (Second Edition)}, R.~Szabo and M.~Bojowald, eds.,
  pp.~507--519.
\newblock Academic Press, Oxford, second edition~ed., 2025.

\bibitem{Gielen:2011dg}
S.~Gielen and D.~Oriti, ``{Discrete and continuum third quantization of
  Gravity},'' in {\em {Quantum Field Theory and Gravity: Conceptual and
  Mathematical Advances in the Search for a Unified Framework}}, pp.~41--64.
\newblock 2012.
\newblock \href{http://arXiv.org/abs/1102.2226}{{\tt arXiv:1102.2226}}.

\bibitem{Carrozza:2020akv}
S.~Carrozza, S.~Gielen, and D.~Oriti, ``{Editorial for the Special Issue
  ''Progress in Group Field Theory and Related Quantum Gravity Formalisms''},''
  {\em Universe} {\bf 6} (2020), no.~1, 19,
  \href{http://arXiv.org/abs/2001.08428}{{\tt arXiv:2001.08428}}.

\bibitem{Oriti:2017ave}
D.~Oriti, {\em {Group field theory and loop quantum gravity.}}, pp.~125--151.
\newblock World Scientific, 2017.

\bibitem{oriti}
D.~Oriti, ``{The group field theory approach to quantum gravity},'' in {\em
  Approaches to Quantum Gravity}.
\newblock Cambridge University Press, Cambridge, 2009.

\end{thebibliography}
\end{document}